\documentclass[preprint,superscriptaddress,nofootinbib%
]{revtex4-1}

\usepackage{amsmath,amssymb,epsfig,color}
\usepackage{booktabs}
\usepackage{bm}

\usepackage{ulem}  % \sout{ text }
\definecolor{darkgreen}{rgb}{0, 0.5, 0.2}

\allowdisplaybreaks[4]
\newcommand{\lsim}{%
\raise0.3ex\hbox{$\;<$\kern-0.75em\raise-1.1ex\hbox{$\sim\;$}}}
\newcommand{\gsim}{%
\raise0.3ex\hbox{$\;>$\kern-0.75em\raise-1.1ex\hbox{$\sim\;$}}}

\renewcommand{\alt}{\lsim}
\renewcommand{\agt}{\gsim}

\newcommand{\non}{\nonumber}
\newcommand{\eq}[1]{(\ref{#1})}
\newcommand{\Eq}[1]{Eq.~(\ref{#1})}

\newcommand{\Eqs}[2]{Eqs.~(\ref{#1})--(\ref{#2})}
\newcommand{\ba}{\begin{eqnarray}}
\newcommand{\ea}{\end{eqnarray}}
\newcommand{\leftB}{\Bigl}
\newcommand{\rightB}{\Bigr}
\newcommand{\ie}{{\it i}.{\it e}.}

\newcommand{\hc}{\mbox{H.c.}}       % in mathmode.

\newcommand{\abs}[1]{\left| #1 \right|}
\newcommand{\wh}[1]{{\widehat #1}}
\newcommand{\wt}[1]{{\widetilde #1}}
\newcommand{\VEV}[1]{\left<#1\right>}
\newcommand{\order}[1]{\mathcal{O}\!\left(#1\right)}
\newcommand{\fun}[1]{\!\left(#1\right)} % fir arguments of a function

\newcommand{\nunits}[1]{\,\mathrm{#1}}  % for numerical values with units

\newcommand{\wb}[1]{%
\vbox{\ialign{##\crcr\hskip 1.0pt\hrulefill\hskip 0.3pt%
\crcr\noalign{\kern-1pt\vskip0.07cm\nointerlineskip}%
$\hfil\displaystyle{#1}\hfil$\crcr}}}

\newcommand{\mxino}{m_{\wt{X}_0}}
\newcommand{\mbino}{m_{\wt{B}}}
\newcommand{\mwino}{m_{\wt{W}^3}}
\renewcommand{\mwino}{m_{\wt{W}}}

\makeatletter
\def\braket<#1|#2|#3>{%
\gdef\@bra{#1}\gdef\@opr{#2}\gdef\@ket{#3}\@braket}
\def\bracket#1{\expandafter\bunri#1\@nil\relax\@braket}
\def\bunri#1|#2|#3\@nil{%
\gdef\@bra{#1}\gdef\@opr{#2}\gdef\@ket{#3}}
\def\@braket{\setbox0=\hbox{$\displaystyle {\@bra}{\@ket}$}
	\dimen0\ht0 \dimen1\dp0
	\setbox0=\hbox{$\displaystyle {\@opr}$}
	\@tempdima=.8\ht0 \@tempdimb=.8\dp0
	\ifdim\@tempdima>\dimen0\relax\dimen0=\@tempdima\fi
	\ifdim\@tempdimb>\dimen1\relax\dimen1=\@tempdimb\fi
	\setbox1=\hbox{\vrule height\dimen0 depth\dimen1 width\z@}
	\def\Strut{\relax\ifmmode\copy1\else\unhcopy1\fi}
	\def\sp@ce{\kern.09em}
	\ifx\@bra\empty\relax
	 \ifx\@opr\empty\relax \left|\Strut\sp@ce{\@ket}\right>
	 \else 
	   \ifx\@ket\empty\relax \left<{\@opr}\right> 
	   \else {\@opr}\left|\Strut\sp@ce{\@ket}\right> \fi\fi
	\else
	 \ifx\@opr\empty\relax
	   \ifx\@ket\empty\relax \left<\Strut{\@bra}\sp@ce\right|
	   \else \left<\Strut{\@bra}\sp@ce\right|\left.
	                         \kern-.22em\Strut{\@ket}\right> \fi
	 \else 
	   \ifx\@ket\empty\relax\left<\Strut{\@bra}\sp@ce\right|{\@opr}
	   \else \left<\Strut{\@bra}\sp@ce\right|{\@opr}
	                \left|\Strut\sp@ce{\@ket}\right> \fi\fi\fi}
\makeatother
\newcommand{\mplanck}{M_{\mathrm{pl}}}           % for reduced planck

\newcommand{\LambdaC}{\Lambda_{\mathrm{cutoff}}} % cutoff for dim.5 op.
\newcommand{\LambdaD}{\Lambda_{\mathrm{D}}}      % cutoff for supersoft op.

\newcommand{\amp}{\mathrm{Amp}}
\renewcommand{\amp}{\mathcal{A}}

\newcommand{\cabbibo}{\varepsilon} 
\newcommand{\suppression}{\mathcal{Z}} % suppression factor of pG mass

\renewcommand{\paragraph}[1]{
\medskip\medskip
\noindent\underline{\hbox{\bf #1}}%${}$\\
}

\begin{document}
%%%%%%%%%%%%%%%%%%%%%%%%%%%%%%%%%%%%%%%%%%%%%%%%%%%%%%%%%%%%%%%%%%%%%%

%%%%%%%%%%%%%%%%%%%%%%%%%%%%%%%%%%%%%%%%%%%%%%%%%%%%%%%%%%%%%%%%%%%%%%
%%%%% declaration for front matter%%%%%%%%%%%%%%%%%%%%%%%%%%%%%%%%%%%%
\title{Neutrino Mass and Proton Decay \\
in a ${U(1)}_R$ Symmetric Model}

\author{Yusuke {\sc Morita}}
\email{morita@muse.sc.niigata-u.ac.jp}
\affiliation{Graduate School of Science and Technology, 
Niigata University,~Niigata, 950-2181, Japan}

\author{Hiroaki {\sc Nakano}}
\email{nakano@muse.sc.niigata-u.ac.jp}
\affiliation{Department of Physics, 
Niigata University,~Niigata, 950-2181, Japan}

\author{Takashi {\sc Shimomura}}
\email{stakashi@muse.sc.niigata-u.ac.jp}
\affiliation{Department of Physics, 
Niigata University,~Niigata, 950-2181, Japan}
\affiliation{Max-Planck-Institut f\"ur Kernphysik, 
Saupfercheckweg 1, D-69117 Heidelberg, Germany}

%\keywords{Supergravity, R-symmetry}

\date{\today}

%\preprint{YITP-11-25}
%\preprint{KUNS-2323}

%%%%%%%%%%%%%%%%%%%%%%%%%%%%%%%%%%%%%%%%%%%%%%%%%%%%%%%%%%%%%%%%%%%%%%
%%%%% typeset front matter (including abstract) %%%%%%%%%%%%%%%%%%%%%%
\begin{abstract}
We study a ${U(1)}_R$ symmetric extenstion of supersymmetric standard model
with supersymmetry breaking in the visible as  well as hidden sectors.
Specifically we study ${U(1)}_R$ breaking effects 
parametrized by the gravitino mass.
A special $R$-charge assignment of right-handed neutrinos allows us 
to have neutrino Yukawa couplings with the $R$-charged Higgs field,
which develops a tiny vacuum expectation value
after the inclusion of $U(1)_R$ symmetry breaking. 
Even with $O(1)$ Yukawa couplings,
a suitable size of Dirac neutrino masses can be generated
if the gravitino mass is very small, 
$m_{3/2}=1\hbox{---}10\,\mathrm{eV}$.
Our flipped $R$-charge assignment also allows 
a new type of dimension five operator that can induce the proton decay.
It turns out that
the proton stability mildly constrains 
the allowed range of the gravitino mass:
Gravitino heavier than $10 \ \mathrm{keV}$ can evade 
the proton decay constraint as well as cosmological ones.
In this case,
the largest neutrino Yukawa coupling is comparable to the electron Yukawa.
We also calculate the mass of the pseudo goldsino
and its mixing to neutralinos,
and briefly discuss its implications in cosmology and Higgs phenomenology.
\end{abstract}
%%%%%%%%%%%%%%%%%%%%%%%%%%%%%%%%%%%%%%%%%%%%%%%%%%%%%%%%%%%%%%%%%%%%%%
\vspace*{-30pt}

%\begin{flushright}
%draft20121216f.tex
%\end{flushright}
%\vspace*{20pt}

\maketitle

\tableofcontents

\newpage

%%%%%%%%%%%%%%%%%%%%%%%%%%%%%%%%%%%%%%%%%%%%%%%%%%%%%%%%%%%%%%%%%%%%%%
%%%%%%%%%%%%%%%%%%%%%%%%%%%%%%%%%%%%%%%%%%%%%%%%%%%%%%%%%%%%%%%%%%%%%%
\section{Introduction}
\label{sec:introduction}
%%%%%%%%%%%%%%%%%%%%%%%%%%%%%%%%%%%%%%%%%%%%%%%%%%%%%%%%%%%%%%%%%%%%%%
%%%%%%%%%%%%%%%%%%%%%%%%%%%%%%%%%%%%%%%%%%%%%%%%%%%%%%%%%%%%%%%%%%%%%%

Weak scale supersymmetry (SUSY) has been an attractive candidate 
for the physics behind the electroweek symmetry breaking (EWSB).
SUSY breaking of order of the weak scale can trigger 
the desired electroweak symmetry breaking
either at tree level or via radiative corrections.
However, null results at LHC for SUSY particle search up to now
would require us to reconsider such picture,
providing a motivation of extending the minimal SUSY Standard Model (MSSM).

$U(1)_R$-symmetric extension of the SUSY standard model is 
an interesting starting point for the physics beyond the MSSM.
A nice feature of $U(1)_R$ symmetry in the matter sector is that
it naturally explains the absence of baryon number violating operators
that would lead to fast proton decay.
See Ref.~\cite{Kribs:2007ac}
for a solution to SUSY flavor problem.
The gauge sector can be made $U(1)_R$ symmetric
%if we introduce the Dirac partner of the MSSM gauginos.
%instead of the usual Majorana mass terms,
%the gauginos have Dirac mass terms, which can be generated
%through so-called supersoft operators from hidden sector SUSY breaking.
%In the gauge sector, $U(1)_R$-symmetry is realized 
if each gaugino has Dirac mass, %term with adjoint chiral multiplet,
instead of the usual Majorana one. 
Such Dirac mass term can be generated from hidden sector SUSY breaking
through a supersoft operator \cite{Fox:2002bu}, 
%\footnote{
%NOTE:See also Ref.~\cite{Chacko:2004mi}.
%},
which induces finite soft scalar masses.
Models with Dirac gaugino have an advantage 
of reducing the degree of fine tuning in the Higgs potential
even when the colored sparticles are 
as heavy as multi TeV \cite{Kribs:2012gx}.
The Higgs sector can also be made $U(1)_R$ symmetric
if we introduce the mirror partner, $R$-partner of the MSSM Higgs doublets.
%so that we have $U(1)_R$-symmetric higgsino mass terms.

The origin of the EWSB can be addressed
in a $U(1)_R$-symmetric manner.
Ref.~\cite{Izawa:2011hi}  proposed\footnote{
One of the motivations behind the construction in Ref.~\cite{Izawa:2011hi}
was to explore the possibility of testing the SUSY breaking mechanism
through the Higgs sector as a portal.
Here we are interested in a simple realization of EWSB
via the visible SUSY breaking.
}
a natural realization of the EWSB
by coupling the $U(1)_R$-symmetric Higgs sector
to visible sector SUSY breaking,
%The constraints from 
in which the supertrace sum rule is avoided
%in the coexistence of the visible and hidden sector SUSY breaking.
by the presence of the hidden sector SUSY breaking.
Such coupling can be used to raise the lightest Higgs mass.
The latter point is important if Dirac gaugino mass terms
are generated by the supersoft operators
which also suppress the tree level $D$-terms,
as was noted in Ref.~\cite{Fox:2002bu}.
See also Ref.~\cite{Benakli:2012cy} and references therein.
Implications in cosmology as well as Higgs search were
also discussed in Ref.~\cite{Bertolini:2011tw},
in which it was pointed out that
the visible SUSY breaking with Majorana gauginos
is cosmologically disfavored.

In the present paper,
we are interested in yet another notable aspect  
of $R$-symmetric extension of SUSY standard model.
The $U(1)_R$ symmetry should be broken in the hidden sector 
for the cosmological constant to be canceled.
If the $R$-symmetry breaking is mediated to the visible sector in a minimal way,
an $R$-partner Higgs field develops a tiny vacuum expectation value (VEV)
characterized by the gravitino mass.
We will show that
such a tiny VEV can be related to the smallness of neutrino masses.
The generation of neutrino mass from SUSY breaking effects 
was studied before in Refs.~\cite{ArkaniHamed:2000bq,MarchRussell:2004uf}.
The generation of (Majorana) neutrino mass from $R$-symmetry breaking 
was also discussed recently in Ref.~\cite{Rehermann:2011ax,Davies:2011js}.
%in the contexts of $R$-symmetric extension of MSSM.
In the present paper,
we will relate the tiny $R$-breaking VEV to Dirac neutrino masses
by flipping the $R$-charge assignment of the right-handed neutrinos.
%We also show how this can be achieved 
%without spoiling so much a nice feature of $U(1)_R$ symmetry 
%of ensuring proton stability.
Our flipped assignment of $U(1)_R$ charges 
allows us to write a desired coupling 
of the right-handed neutrino to the $R$-partner Higgs doublet;
at the same time, it also allows a new type of dimension five operator
that makes the proton unstable.
It turns out that
the constraint from the proton stability 
is milder than the corresponding constraints in the MSSM.
Thus in our flipped $U(1)_R$ model,
tiny neutrino masses can be generated 
without spoiling so much a nice feature of $U(1)_R$ symmetry 
of ensuring proton stability.
%of suppressing proton decay.

The present setup of $U(1)_R$-symmetric model has 
potentially interesting implications on Higgs phenomenology.
Since we consider the model of visible SUSY breaking 
and also assume that Dirac gaugino masses are induced from 
another SUSY breaking in a hidden sector,
there appears a physical pseudo goldstino state 
$\zeta$ \cite{Cheung:2010mc},
which directly couples to the Higgs sector.
As discussed in Ref.~\cite{Bertolini:2011tw},
such pseudo goldstino can affect the Higgs decay modes.
%, and the constraints from cosmology are also important.
We perform a similar analysis in our $U(1)_R$ symmetric setup
and examine to some details the mass and the mixing of the pseudo goldstino.
It turns out that
although the pseudo goldstino can get a mass due to $R$-breaking effect,
its mass $m_\zeta$ is quite suppressed compared to the gravitino mass $m_{3/2}$.
This is contrasted to the `sequestered' case 
in which $m_\zeta$ is twice as large as $m_{3/2}$ 
\cite{Cheung:2010mc,Argurio:2011hs}.
Moreover,
%\red{
the mixings of the pseudo goldstino to the MSSM Higgsinos
and gauginos
are highly suppressed due to the softly broken $U(1)_R$ symmetry.
%}

Another important constraint on the present model 
comes from cosmology;
the light gravitino is constrained rather severely
not to disturb
the successful Big-Bang Nucleosynthesis (BBN),
the structure formation of galaxies
and the cosmic microwave background (CMB) radiations,
and not to be produced too much in the early universe.
As we shall see later,
the allowed range of the gravitino mass is given by\footnote{
See Ref.~\cite{Shirai:2010rr} and references therein.
See also Sec.~\ref{sec:implications} for a brief discussion.
}
\begin{align}
m_{3/2} \alt 16 \nunits{eV}
    \ , \qquad
10\nunits{keV} \alt m_{3/2} %\alt 10\,\hbox{--}\,100 \ \mathrm{MeV} 
    \ .
\label{eq:gravitino:range}
\end{align}
We will take account of these limits in the following analysis.
We will also make a brief comment
on the limit on decaying dark matter scenario
from the diffused gamma-ray line search.

The paper is organized as follows.
In the next section,
we present our model first in the $R$-symmetric limit,
and we then include $R$-breaking effects in a minimal way
that are parameterized by the gravitino mass.
Throughout the present paper,
we assume the gravitino mass to be much smaller than the weak scale,
although we do not elucidate possible origins of hidden sector SUSY breaking.
In section three,
we show how the small VEVs of the $R$-charged Higgses
can be used to explain the smallness of the neutrino mass.
We also discuss the constraint from the proton decay
in our flipped $U(1)_R$ model for neutrino mass.
In section four,
we calculate the pseudo goldstino mass
and the mixings to neutralinos
under the assumption of minimal $U(1)_R$ breaking.
Specifically we show how the mass of the pseudo goldstino
is suppressed compared with the gravitino
in the combined model of Dirac gaugino and visible SUSY breaking.
%The mixing between the the pseudo goldstino and neutralinos
%can be used to
We then apply the results
to briefly discuss implications of the present model
on Higgs phenomenology and cosmology.
The final section is devoted to our conclusion and discussion.
We also add two appendices
concerning the analysis on baryon and/or lepton number violating operators
and a possible extension of the model
in which the constraint from proton stability
will be relaxed via $U(1)$ flavor symmetry.

For simplicity, we will often refer to $U(1)_R$ symmetry 
as ``$R$ symmetry'', if no confusion is expected.

%\newpage
%%%%%%%%%%%%%%%%%%%%%%%%%%%%%%%%%%%%%%%%%%%%%%%%%%%%%%%%%%%%%%%%%%%%%%
%%%%%%%%%%%%%%%%%%%%%%%%%%%%%%%%%%%%%%%%%%%%%%%%%%%%%%%%%%%%%%%%%%%%%%
%\section{visible SUSY breaking with broken $R$-symmetry}
\section{$R$-Symmetric Model with Visible SUSY Breaking}
\label{sec:model}
%%%%%%%%%%%%%%%%%%%%%%%%%%%%%%%%%%%%%%%%%%%%%%%%%%%%%%%%%%%%%%%%%%%%%%
%%%%%%%%%%%%%%%%%%%%%%%%%%%%%%%%%%%%%%%%%%%%%%%%%%%%%%%%%%%%%%%%%%%%%%

In this section,
we present an extension of the MSSM,
in which $U(1)_R$ symmetry is realized
by combining the model of Dirac gauginos \cite{Fox:2002bu}
with the model of visible SUSY breaking of Ref.~\cite{Izawa:2011hi}.
We then discuss the effects of the minimal  $R$-symmetry breaking
induced from the coupling to supergravity.

%%%%%%%%%%%%%%%%%%%%%%%%%%%%%%%%%%%%%%%%%%%%%%%%%%%%%%%%%%%%%%%%%%%%%%
\subsection{$U(1)_R$ Symmetric Model}
%%%%%%%%%%%%%%%%%%%%%%%%%%%%%%%%%%%%%%%%%%%%%%%%%%%%%%%%%%%%%%%%%%%%%%

The $U(1)_R$ symmetry forbids
the usual Majorana gaugino mass terms
as well as the higgsino mass term, $\mu H_u H_d$, in the MSSM.
To realize $U(1)_R$ symmetry, we extend the MSSM by introducing
the $R$-partners of the gauginos and those of the higgsinos.

%\paragraph{Gauge sector}

First let us briefly discuss the gauge sector.
Let $a=3,2,1$ parameterize each gauge group of 
$SU(3)_C\times SU(2)_L \times U(1)_Y$.
{}For each gauge group $G_a$,
we introduce an adjoint chiral multiplet $A_a$ 
which contains an $R$-partner $\chi_a$ of the gaugino $\lambda_a$.
Since the gaugino $\lambda_a$ has $R$-charge $+1$,
its partner $\chi_a$ has to have $R$-charge $-1$
so that the adjoint chiral superfield $A_a$ has $R=0$.
Accordingly, 
%instead of the usual Majorana gaugino mass term
%which breaks the $U(1)_R$ symmetry,
we have a Dirac type mass term
\begin{align}
\mathcal{L}_{\mathrm{gaugino}}
\ ={}-\sum_{a=1,2,3} m_{a} \lambda_a \chi_a + \hc
      \ .
%\ ={}-m_B \tilde\chi_B \tilde{B}- m_W \tilde\chi_W \tilde{W} + \hc \ .
\end{align}
The Dirac gaugino mass term can be generated 
through the ``supersoft" operator \cite{Fox:2002bu}
\begin{equation}
\mathcal{L}_{\mathrm{supersoft}}
\ =\ \sum_{a=1,2,3} \int d^2 \theta\,\sqrt{2}  
     \frac{{W'}^\alpha W_\alpha^a A_a}{\LambdaD}
    + \hc
 \ , 
\label{eq:supersoft}
\end{equation}
where $\LambdaD$ is a messenger scale,
$W_\alpha^a$ is the MSSM gauge field strength,
$W'_\alpha$ is a hidden-sector gauge field strength 
which acquires a nonzero $D$-term,
$\VEV{W'_\alpha}=\theta_\alpha\VEV{D'}$
so that $m_a=\VEV{D'}/\LambdaD$.
In the present work, 
we assume that a suitable size of masses are generated
although we do not elucidate the hidden sector dynamics.

%%%%%%%%%%%%%%%%%%%%%%%%%%%%%%%%%%%%%%%%%%%%%%%%%%%%%%%%%%%%%%%%%%%%%%
%\input{table.tex}
\begin{table}[tdp]
\begin{center}
\begin{tabular}{c|ccccc|ccc}
 & \makebox[12mm]{ $X_0$ }
 & \makebox[10mm]{ $X_u$ }
 & \makebox[10mm]{ $X_d$ }
 & \makebox[10mm]{ $H_u$ }
 & \makebox[12mm]{ $H_d$ }
 & \makebox[12mm]{ $L$   }
 & \makebox[10mm]{ $E$   }
 & \makebox[10mm]{ $N$   } \\ \hline \hline
 $SU(2)_L$ 
 & $\mathbf 1$ 
 & $\mathbf 2$ & $\mathbf 2$ 
 & $\mathbf 2$ & $\mathbf 2$ 
 & $\mathbf 2$ & $\mathbf 1$
 & $\mathbf 1$ \\
$U(1)_Y$ 
 &  $0$ & $+1/2$ & $-1/2$ & $+1/2$  & $-1/2$ & $-1/2$ & $+1$ &  $0$ \\ \hline
$U(1)_R$
 & $+2$ &   $+2$ &   $+2$ &    $0$  &    $0$ &   $+1$ & $+1$ & $-1$ \\
\end{tabular}
\end{center}
\caption{
The charge assignment of the Higgs and lepton fields 
under the EW symmetry and $U(1)_{R}$ symmetry.
The $R$-charges are those for left-handed chiral superfields.
The MSSM Higgs doublets are $R$-neutral while 
their $R$-partners and the singlet $X_0$ have $R$-charge $+2$.
All the quarks and leptons have $R$-charge $+1$,
except that the right-handed neutrinos have $R$-charge $-1$.
The implications of this flipped assignment 
will be discussed in Sec.~\ref{sec:flipped}.
Note also that
the adjoint chiral multplets $A_a$ are $R$-neutral
so that their fermionic components have $R$-charge $-1$.
}
\label{tab:higgs}
\end{table}
%%%%%%%%%%%%%%%%%%%%%%%%%%%%%%%%%%%%%%%%%%%%%%%%%%%%%%%%%%%%%%%%%%%%%%

%\paragraph{Higgs sector}

Next, we turn to the Higgs sector.
Following the Ref.~\cite{Izawa:2011hi}, we consider the superpotential
\begin{align}
W_{\mathrm{Higgs}}
 &= X_0 \left(f+\lambda H_u H_d \right) + \mu_1 X_d
    H_u + \mu_2 X_u H_d
    \ .
\label{eq:visibleW}
\end{align}
The gauge and $R$-charge assignments are shown in Table.~\ref{tab:higgs}.
The superfields $X_d$ and $X_u$ with $R$-charge $2$
are the mirror partners of the MSSM Higgs fields $H_u$ and $H_d$,
with the supersymmetric masses $\mu_1$ and $\mu_2$,
respectively:
The $SU(2)\times U(1)$ singlet field $X_0$ also has $R$-charge $2$.
The dimension two parameter $f$
is the source of the visible SUSY breaking
while the dimensionless coupling $\lambda$ plays important roles
not only for triggering the EWSB 
but also for generating the quartic coupling of the MSSM Higgs scalars.

With the $R$-symmetric superpotential \eq{eq:visibleW},
supersymmetry is spontaneously broken
in accordance with the general argument\footnote{
There exists no supersymmetric vacuum 
when the number of the fields with $R$-charge $2$
is larger than the number of the fields with $R$-charge $0$.
%when there are more fields with $R$-charge $2$
%compared with the number of fields of $R$-charge 0.
See Ref.~\cite{Ray:2007wq} and references therein.
}.
Moreover,
such visible SUSY breaking triggers the correct EWSB
if the coupling $\lambda$ is sufficiently large.
A possible origin of the dimensionful parameters $f$ and $\mu_{1,2}$
was suggested in Ref.~\cite{Izawa:2011hi}.
Here we just assume that
the scale of these parameters is around the weak scale;
%\eg, 
we will take $\mu_{u,d}=300\nunits{GeV}$ 
in our analysis in the next section.

%\paragraph{Note}

We note that
the adjoint chiral multiplets $A_{a=2,1}$ of $SU(2)\times U(1)$
can have a $U(1)_R$-symmetric superpotential interactions 
with the Higgs fields $H_{u,d}$ and $X_{d,u}$.
For simplicity, however, we do not include them in our analysis.
%\begin{align}
%W_{\mathrm{adj}}
% &=  X_u \cdot A_2 H_d + X_d \cdot A_2 H_u
%    + X_u  A_1 H_d + X_d  A_1 H_u 
%\non\\
% &=  \sum_{a=1,2}
%      \left(X_u\right)^\alpha
%      \left[
%        A_2^a \left(\tau^a\right)_{\alpha}{}^\beta 
%      + A_1 \delta_{\alpha}^\beta
%      \right]
%      \left(H_d\right)_\beta
%     +\sum_{a=1,2}
%      \left(X_d\right)^\alpha
%      \left[
%        A_2^a \left(\tau^a\right)_{\alpha}{}^\beta 
%      + A_1 \delta_{\alpha}^\beta
%      \right]
%      \left(H_u\right)_\beta
%    \ , \qquad
%\end{align}
More importantly, we do not include the following mixing term
between the singlet $X_0$ and the $U(1)_Y$ ``adjoint'' multiplet $A_1$.
\begin{align}
W_{\mathrm{mix}} \ =\ \mu_0 X_0 A_1 \ .
\end{align}
This term is very dangerous since it could cancel, if present,
the linear term in the superpotential \eqref{eq:visibleW}.
The absence of such term can be justified if 
$U(1)_Y$ is embedded into a simple gauge group at high energy.
%along the line of Ref.\cite{Fox:2002bu}.
See Ref.~\cite{Fox:2002bu} for a similar discussion
about the absence of the kinetic mixing of the hidden $U(1)'$ and $U(1)_Y$.

%[The possible embedding into a (semi-)simple gauge group
%were discussed in Ref.~\cite{Fox:2002bu}....]

%%%%%%%%%%%%%%%%%%%%%%%%%%%%%%%%%%%%%%%%%%%%%%%%%%%%%%%%%%%%%%%%%%%%%%
\subsection{Minimal $U(1)_R$ Symmetry Breaking}
%%%%%%%%%%%%%%%%%%%%%%%%%%%%%%%%%%%%%%%%%%%%%%%%%%%%%%%%%%%%%%%%%%%%%%

Next we discuss $U(1)_R$ symmetry breaking. %and its consequences.
Once the model is coupled to supergravity,
the $U(1)_R$ symmetry is necessarily broken
in order that the cosmological constant can be adjusted to zero.
Our basic assumption here is that
such breaking of $R$ symmetry is mediated to the visible sector
in a minimal way; % of anomaly mediation.
under such assumption of ``minimal $R$-breaking mediation'',
the interaction Lagrangian is given by\footnote{
Here we took a modified gauge fixing of Ref.~\cite{Cheung:2011jp}.
}
\begin{align}
\mathcal{L}_{\mathrm{eff}}
 &= \int{d^2 \theta}\,\phi^3W%_{\mathrm{Higgs}}
   +\hc
    \ , \qquad 
\phi = 1 + \theta^2 m_{3/2}
    \ , 
\end{align}
where $\phi$ is the so-called conformal compensator
and $m_{3/2}$ is the mass of the gravitino.
The conformal compensator $\phi$ can be absorbed
by rescaling the chiral superfields,
$\phi \Phi_i \rightarrow \Phi_i$,
where %$\Phi_i=H_u,H_d,X_u,X_d$ or $X_0$. 
$\Phi_i$ represents all the chiral superfields in the theory.
Classically such rescaling has no effect
on cubic terms in superpotential,
whereas terms of dimensions less than four are affected.
%With the same notation for scalar components as 
%the corresponding superfields,
In the present case, the interaction Lagrangian takes the form
\begin{align}
\mathcal{L}_\text{eff}
 &= \int{d^2 \theta}\,W%_{\mathrm{Higgs}} 
   + m_{3/2}G(\Phi_i) + \hc \ , 
\label{eq:lag_eff} \\
G(\Phi_i)%,\mu_1,\mu_2)
 &\equiv 2fX_0+ \mu_1 X_d H_u +\mu_2 X_u H_d
     \ ,
\label{eq:G}
\end{align}
where 
the fields in $G(\Phi_i)$ are 
the scalar components of the corresponding superfields
$\Phi_i=H_u,H_d,X_u,X_d$ or $X_0$. 
%The first term in Eq.~\eqref{eq:lag_eff} is $R$-symmetric
Terms %in Eq.~\eqref{eq:lag_eff},
proportional to the gravitino mass
represent $R$-breaking interactions,
and are small if the gravitino mass is small;
we assume throughout the present paper that
the gravitino mass is much smaller than the weak scale.
%$m_{3/2}\alt 10\nunits{meV}$ or smaller.]
As we will see shortly, however,
these interactions induce a slight shift of the vacuum
since they contain a tadpole term of $X_0$ 
and also those of $X_{u,d}$ after the EWSB.
Such shift of the VEVs will play important roles
when we discuss the generation of Dirac neutrino masses
(in \S{\ref{sec:flipped}})
and properties of the pseudo goldstino (in \S{\ref{sec:goldstino}}).

%%%%%%%%%%%%%%%%%%%%%%%%%%%%%%%%%%%%%%%%%%%%%%%%%%%%%%%%%%%%%%%%%%%%%%
\subsection{The EWSB Vacuum}
%%%%%%%%%%%%%%%%%%%%%%%%%%%%%%%%%%%%%%%%%%%%%%%%%%%%%%%%%%%%%%%%%%%%%%

We now analyze the scalar potential in order to find 
the shift of the VEVs induced by the $R$-breaking effects.
Here we assume that the K\"ahler potential is canonical.
%[Later we will discuss the goldstino mass matrix
%with non-canonical K\"ahler potential for $X_0$.]
The scalar potential involving the neutral Higgs fields 
$H_{u,d}^0$ and their mirror partners $X_{u,d}^0$ 
and the singlet $X_0$ is given by 
\begin{align}
%V\!\left(\Phi_i,\Phi_i^\dagger\right) 
V(\Phi_i,\Phi_i^\dagger) 
 &= V_0(\Phi_i,\Phi_i^\dagger) 
    -\left\{ m_{3/2}G\!\left(\Phi_i\right) 
            +\hbox{H.c.}
     \right\} \ , 
\label{eq:potential} 
\end{align}
with the $R$-symmetric part $V_0$ given by 
\begin{align}
V_{0}
  &=  
     \left| f - \lambda H_u^0 H_d^0 \right|^2
   + \left| \lambda X_0 H_d^0 - \mu_1 X_d^0 \right|^2 
   + \left| \lambda X_0 H_u^0 + \mu_2 X_u^0 \right|^2  
\nonumber \\
  &\quad
    +\frac{1}{8} g^2 \left(
       \left|H_u^0\right|^2 - \left|H_d^0\right|^2 
       +\left|X_u^0\right|^2 - \left| X_d^0 \right|^2 
     \right)^2 
\nonumber \\
  &\quad 
    + \wb{m}_{H_u}^2 \left| H_u^0 \right|^2  
    + \wb{m}_{H_d}^2 \left| H_d^0 \right|^2 
    - \left( b H_u^0 H_d^0 + h.c. \right)
\nonumber \\
  &\quad 
    + m_{X_0}^2 \left|X_0  \right|^2 
    + m_{X_u}^2 \left|X_u^0\right|^2 
    + m_{X_d}^2 \left|X_d^0\right|^2. 
\label{scapot}
\end{align}
The first and the second lines represent 
the $F$-term and the $D$-term potentials, respectively.\footnote{
As noted in Ref.~\cite{Fox:2002bu}
the $SU(2)\times{}U(1)$ $D$-terms are absent
in the supersoft limit.
If the adjoint scalars get soft scalar masses
through $R$-invariant mediation of SUSY breaking,
then $D$ terms does not decouple completely.
Our results in the present paper are not affected 
whether the $D$-terms decouple or not.
}
The coupling constant $g^2$ is $g_1^2 + g_2^2$ 
where $g_1$ and $g_2$ are the  gauge coupling constants 
of $U(1)$ and $SU(2)$ symmetries, respectively. 
The remaining are soft SUSY breaking terms
except that
the masses of the MSSM Higgs fields, 
$\wb{m}_{H_u}^2$ and $\wb{m}_{H_d}^2$, 
are the sums of soft masses and the supersymmetric masses.
We suppose that
these soft terms are induced radiatively from the supersoft operators,
or mediated from the hidden sector in an $R$-invariant way.
We also assume that
%For simplicity, we assume that 
all dimensionful parameters are around or just above the weak scale 
whereas the gravitino mass is much smaller. % than the EWSB scale. 
In the following, we consider the case in which
$m_{X_0}^2$ and $m_{X_{u,d}}^2$ are positive so that 
$X_0$ and $X_{u,d}$ do not develop VEVs 
in the $R$-symmetric limit.
%with vanishing gravitino mass.

The vacuum with EWSB can be found by solving the stationary conditions
\begin{align}
0\ =\ \frac{\partial V}{\partial \Phi_i}
 \ =\ \frac{\partial V_{0}}{\partial \Phi_i}
     - m_{3/2} \frac{\partial G\!\left(\Phi_i\right)}{\partial \Phi_i}
    \ .
\label{eq:vac-cond}
\end{align}
We solve these equations perturbatively 
with respect to a small gravitino mass. 
In the $R$-symmetric limit, \ie, $m_{3/2}\rightarrow0$,
the solution %to the stationary conditions 
was given in Ref.~\cite{Izawa:2011hi},
which we denote by $\VEV{\Phi_i}_{0}$.
Note that the unbroken $U(1)_R$ symmetry implies 
$\VEV{X_0}_0=\VEV{X_u}_0=\VEV{X_d}_0=0$.

Now, with the $R$-breaking terms,
the solution to Eq.~\eqref{eq:vac-cond} takes the form
\begin{align}
\VEV{\Phi_i}
\ =\ \VEV{\Phi_i}_{0} + \kappa_i m_{3/2}
    +\order{m^2_{3/2}} \ , 
\label{eq:vac-sol}
\end{align}
where the coefficient $\kappa_i$ is found to be
\begin{align}
\kappa_i 
\ =\ M^{-1}_{ij} 
       \left. \frac{\partial G}{\partial \Phi_j} 
       \right|_{\Phi = \VEV{\Phi}_0} \ , \qquad
\label{eq:kappa} 
M_{ij}
 \ \equiv\  
       \left. 
         \frac{\partial^2 V_{0}}{\partial \Phi_i \partial \Phi_j}
       \right|_{\Phi =\VEV{\Phi}_0} 
       \ .
%%\label{eq:Mij}
\end{align}
Explicitly we obtain, for the $R$-charged Higgs fields,
\begin{align}
\VEV{X_0}
 &= %m_{3/2}\,
    \frac{2 f}{m^2_{X_0} + \lambda^2v^2}
    \,m_{3/2}
    \ , 
\label{eq:vev-x0} \\
\VEV{X^0_d}
 &= %m_{3/2}\,
    \frac{\mu_1 v\sin\beta
          }{m^2_{X_d} + \mu^2_1 + \left(m^2_Z/2\right)\cos2\beta}
    \,m_{3/2} 
    \ , 
\label{eq:vev-x1}\\
\VEV{X^0_u}
 &= %m_{3/2}\,
    \frac{{}-\mu_2 v\cos\beta 
          }{m^2_{X_u} + \mu^2_2 - \left(m^2_Z/2\right)\cos2\beta}
    \,m_{3/2} 
    \ , 
\label{eq:vev-x2}
\end{align}
where $m_Z^2 = \left(g_1^2+g_2^2\right) v^2 /2$ 
is the $Z$ boson mass.\footnote{
Note that
the VEVs of the MSSM Higgs fields also receive $\order{m_{3/2}}$ shifts,
which are negligibly small for a small gravitino mass.
}
We see that 
the VEV of $X_0$ is proportional to $f$ that represents 
the scale of the visible SUSY breaking,
while the VEVs of $X_{u,d}$ are proportional to 
their supersymmetric masses and their partner's VEVs. 
These proportionality can be understood if one notices that
replacing the Higgs fields by their VEVs in \Eq{eq:G}
generates the tadpole term for each of $X_{0,u,d}$,
\begin{align}
G\fun{X_0,X_u^0,X_d^0}
\ =\ 2fX_0
    +\mu_1\VEV{H_u^0}X_d^0
    -\mu_2\VEV{H_d^0}X_u^0 \ .
\label{eq:tadpole}
\end{align}
Of course,
the VEVs of these $R$-charged fields
are all proportional to the gravitino mass\footnote{
Precisely speaking,
the VEVs $\VEV{X_{0,u,d}}$ have $R$-charge $+2$
and should be proportional to $m_{3/2}^*$
since one can regard the gravitino mass 
as having $R$-charge $-2$
in the sense of spurion analysis.
}
which parameterizes the $U(1)_R$-symmetry breaking.

% One can see from Eq.~\eqref{eq:vev-x0} that 
%the VEV of $X_0$ is proportional to $f$ that represents 
%the scale of the visible SUSY breaking.  
%On the other hand, from Eqs.~\eqref{eq:vev-x1} and \eqref{eq:vev-x2}, 
%the VEVs of $X_{u,d}$ are proportional to their supersymmetric masses 
%and the VEVs of the corresponding Higgs fields. 
%These proportionality can be understood %from Eq.~\eqref{eq:potential} 
%if one 
%that the tadpole terms are generated 
%by replacing the Higgs fields by their VEVs.

%It is also noted that 
%the fields with non-zero $R$-charges have VEVs proportional to $m_{2/3}$ 
%which characterizes the scale of $R$-symmetry breaking. 
%[See \eqref{eq:vev-x0}-\eqref{eq:vev-x2}.]
%The new contributions to the Higgs fields can be obtained 
%from Eq.~\eqref{eq:kappa}. However, these are negligibly small 
%compared to the $R$-symmetric solutions due to tiny gravitino mass. 

%%%%%%%%%%%%%%%%%%%%%%%%%%%%%%%%%%%%%%%%%%%%%%%%%%%%%%%%%%%%%%%%%%%%%%
%%%%%%%%%%%%%%%%%%%%%%%%%%%%%%%%%%%%%%%%%%%%%%%%%%%%%%%%%%%%%%%%%%%%%%
%\section{Neutrino Mass from SUGRA}
\section{%
Neutrino Mass and Proton Decay in Flipped $U(1)_R$ Model}
\label{sec:flipped}
%%%%%%%%%%%%%%%%%%%%%%%%%%%%%%%%%%%%%%%%%%%%%%%%%%%%%%%%%%%%%%%%%%%%%%
%%%%%%%%%%%%%%%%%%%%%%%%%%%%%%%%%%%%%%%%%%%%%%%%%%%%%%%%%%%%%%%%%%%%%%

In this section, we show that 
the neutrinos can acquire tiny masses
via the supergravity-induced effect of  $U(1)_R$-symmetry breaking.
We introduce three right-handed neutrino multiplets $N^c_i$ ($i=1,2,3$) 
and assign the $R$-charge $-1$ to them
as shown in Table.~\ref{tab:higgs}.
That is, we flip the $R$-charge of the right-handed neutrinos.
%This is the central assumption whose implications 
%we shall discuss in the present paper.
Under this flipped assignment,
we first argue that the neutrinos can be Dirac particles,
and that the smallness of neutrino masses are related to
the smallness of $U(1)_R$-symmetry breaking 
parameterized by the gravitino mass.
Then, we discuss the proton decay 
induced by a new type of dimension five operator 
involving the right-handed neutrinos.

%%%%%%%%%%%%%%%%%%%%%%%%%%%%%%%%%%%%%%%%%%%%%%%%%%%%%%%%%%%%%%%%%%%%%%
\subsection{Neutrino Mass in Flipped $U(1)_R$ Model}
\label{sec:neutrino}
%%%%%%%%%%%%%%%%%%%%%%%%%%%%%%%%%%%%%%%%%%%%%%%%%%%%%%%%%%%%%%%%%%%%%%

%\begin{align}
%W
% \ = \ y_\nu^{ij} N^c_i L_j X_u 
%     \ ,
%\end{align}
%%Here we omit flavor indices for simplicity.
%Since we are most interested in the heaviest neutrino,
%we will suppress the flavor indices hereafter.

Under our flipped $U(1)_R$ assignment,
the right-handed neutrino can have a Yukawa-type interaction 
with the mirror Higgs field $X_u$, instead of the MSSM Higgs $H_u$.
At the renormalizable level
the matter superpotential is given by
\begin{align}
W = y_U^{ij} Q_i U^c_j H_u + y_D^{ij} Q_i D^c_j H_d 
   +y_E^{ij} E^c_i L_j H_d + y_N^{ij} N^c_i L_j X_u
    \ ,
\label{eq:yukawa}
\end{align}
where $y_{U,D,E,N}$ are $3 \times 3$ Yukawa coupling matrices.

The Majorana mass term for the right-handed neutrinos 
is forbidden by our $R$-charge assignment. 
The symmetry allows the Weinberg operator
\begin{align}
W_{\Delta{L}=2}
\ ={}-\frac{C_{\Delta{L}=2}^{ij}}{\LambdaC}
      \left(L_i H_u\right)\left(L_j H_u\right)
      \ ,
\label{eq:weinbergop}
\end{align}
which gives a sub-dominant contribution to the neutrino mass matrix
%if the cutoff $\LambdaC$ is as large as the Planck scale.
if we assume the cutoff $\LambdaC$ 
to be larger than $10^{12}$--$10^{13}\nunits{GeV}$.
Therefore the neutrinos are almost Dirac particles in our model.

By replacing the mirror Higgs field with its VEV \eqref{eq:vev-x2}, 
we obtain the Dirac mass term for neutrinos as
\begin{align}
\mathcal{L}_{\nu\,\mathrm{mass}}
\ ={} -m_\nu \nu_R^c \nu_L + \hc
      \ , \qquad
m_\nu^{ij}
\ =\ y_N^{ij} \VEV{X_u^0} \ .
\end{align}
Since we are most interested in the heaviest neutrino,
we will suppress the flavor indices
%\footnote{
%NOTE: 
%$\nu_{R}^j=N^c_i\left(U_{\text{MNS}}\right)^{ij}$
%in a flavor basis in which the charged fermion mass matrix is diagonal.
%}
hereafter and denote 
the heaviest eigenvalue and the corresponding Yukawa coupling 
by $m_{\nu}$ and $y_{\nu}$, respectively.
With this understanding,
we write
\begin{align}
m_\nu 
\ =\ y_\nu \VEV{X_u^0}
\ =\ y_\nu m_{3/2}
     \left[
       \frac{{}-\mu_2 v\cos\beta
             }{m^2_{X_u}+\mu^2_2- \left(m^2_Z/2\right)\cos2\beta}
     \right]
     \ .
\label{numass}
\end{align}

A notable feature in the present model is that
the size of the neutrino mass is set
by that of the gravitino mass $m_{3/2}$
when all other dimensionful parameters are of the same order.
Accordingly
the neutrino Yukawa coupling can be of $\order{1}$
if the gravitino is as light as $10\nunits{eV}$.
In fact, as we mensioned in \Eq{eq:gravitino:range},
the mass of the gravitino is constrained cosmologically;
%as we shall review in Sec.~\ref{sec:cosmology},
%%from Big-Bang Nucleosynthesis (BBN), 
%%structure formation of Galaxy \red{and gamma-ray line}:
%%\cite{Ichikawa:2009ir, Viel:2005qj, Boyarsky:2008xj},
%the allowed range is given by 
%%\cite{Viel:2005qj,Boyarsky:2008xj,Endo:2007sz,Ichikawa:2009ir}
\begin{align}
m_{3/2} \alt 16 \nunits{eV}
    \ , \qquad
10\nunits{keV} \alt m_{3/2} 
%    \hbox{\red{$\alt 10 \ \nunits{MeV}$}}
    \ .
%\label{eq:gravitino:range}
\non
\end{align}
%the gravitino mass should be either $\lsim 16\nunits{eV}$ 
%or $10\nunits{keV} \alt m_{3/2} \alt 10\nunits{MeV}$ 
%so that the gravitino does not disturb the standard thermal history 
%of the Universe \cite{Viel:2005qj,Boyarsky:2008xj,Endo:2007sz}.
In the case of lighter gravitino, 
a tiny neutrino mass can be obtained with a large Yukawa coupling.
Even in the case of heavier gravitino,
the neutrino Yukawa can be comparable to the electron Yukawa.
%\red{
%We shall defer the further discussion of cosmological constraints
%to Sec.~\ref{sec:cosmology}.
%}

%%%%%%%%%%%%%%%%%%%%%%%%%%%%%%%%%%%%%%%%%%%%%%%%%%%%%%%%%%%%%%%%%%%%%%
\begin{figure}[tb]
\begin{minipage}{0.95\linewidth}
\includegraphics[width=0.45\linewidth]{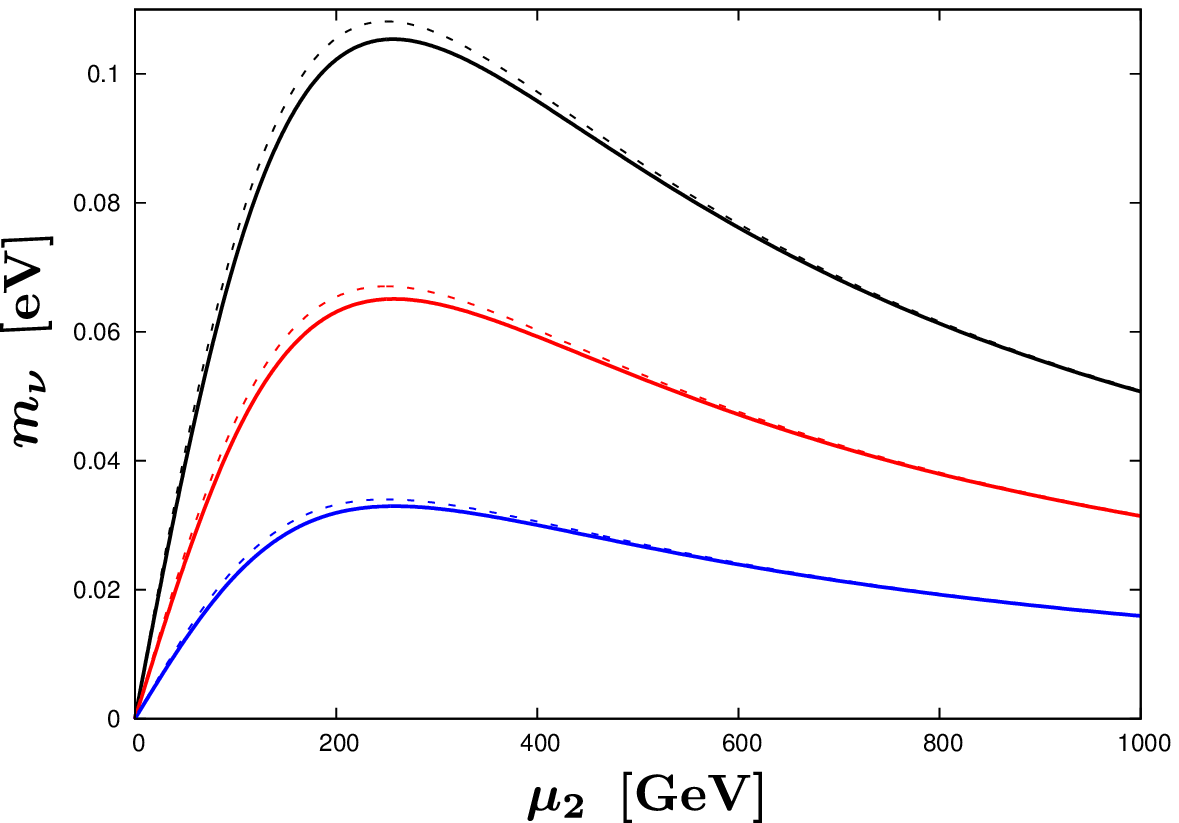} 
 \hfill
\includegraphics[width=0.45\linewidth]{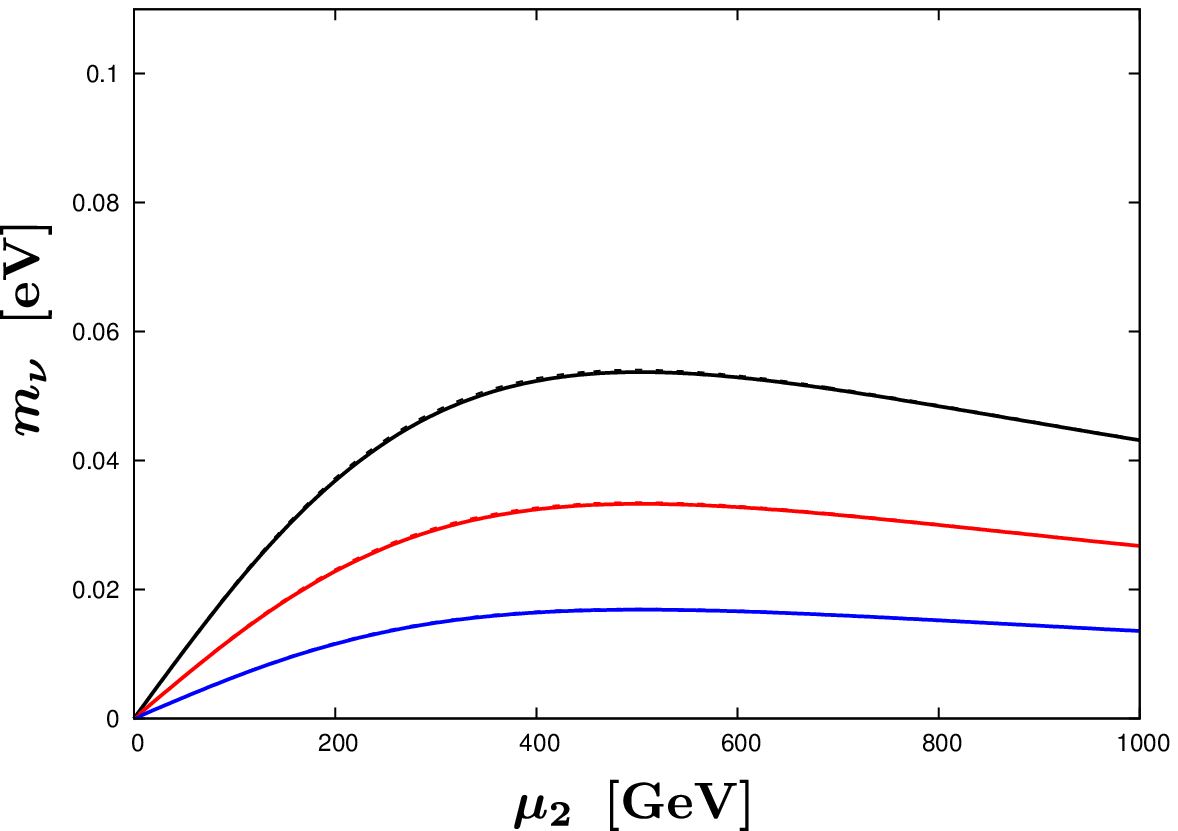} 
\end{minipage}
\caption{
Left:
The heaviest neutrino mass 
as a function of the Higgsino mass parameter $\mu_2$,
with $m_{X_d}=250\nunits{GeV}$ (Left) and $500\nunits{GeV}$ (Right).
In each figure,
the solid and dotted lines correspond to 
the case with and without the $D$-term contribution
in the denominator in \Eq{numass}.
Three lines correspond to $\tan\beta=3$ (top),
$\tan\beta=5$ (middle), and $\tan\beta=10$ (bottom), respectively.
}
\label{fig:neutrinomass}
\end{figure}
%%%%%%%%%%%%%%%%%%%%%%%%%%%%%%%%%%%%%%%%%%%%%%%%%%%%%%%%%%%%%%%%%%%%%%

Fig.~\ref{fig:neutrinomass} shows how the generated neutrino mass
depends on the Higgs mass parameters,
for a fixed value of $m_{3/2}=10\nunits{eV}$ and $y_\nu=0.1$.
We see that
the neutrino mass is maximized for $\mu_2\sim m_{X_u}$.
We also note that
the $D$-term contribution 
in the denominator in \Eq{numass}
can safely be neglected
when the Higgs mass parameters are larger than $m_Z$.
Accordingly the neutrino mass
is proportional to $\cos\beta$ instead of $\sin\beta$.

%\red{(What value of $m_{3/2}$ is used in the Figure?)}
%\blue{$\leftarrow$ ????}

Some remarks are in order.
First,
our $U(1)_R$-charge assignment
does not induce the mixed anomalies with the SM gauge groups
since the right-handed neutrinos are SM singlets.
$U(1)_R$ anomalies can be canceled 
by introducing other singlets with positive $U(1)_R$-charges,
without changing our results below.
%[\red{Mass term?}]
Second,
our model can be regarded as a realization of 
``neutrinophilic Higgs'' idea:
the original idea was proposed
in non-SUSY context
in Refs.~\cite{Ma:2000cc,Ma:2001mr} 
in which a softly broken $Z_2$ symmetry is the source of the tiny VEV.
%[See also \cite{Gabriel:2006ns}?]
See also Ref.~\cite{PhysRevD.80.095008}
for a softly broken global $U(1)$ case.
Quantum stability was discussed 
in Ref.\cite{PhysRevD.85.055002,Haba201198ai}.
The point in our model is that
the size of the ``neutrinophilic'' Higgs VEV is related
to that of the gravitino mass,
and we can address its implications
in cosmology and Higgs phenomenology,
in addition to the proton stability as we shall discuss shortly.
We also note that 
%many implications were discussed, \eg, in
many other topics were discussed in literature,
such as muon $g-2$ and lepton flavor violations 
\cite{Ma:2001mr,PhysRevD.80.095008}, %Ma and Raidal
low-scale leptogenesis \cite{Haba:2011ra},
dark matter and cosmology \cite{PhysRevD.86.043515}, %gamma-ray line
dark energy \cite{Wang:2006jy},
and supernova neutrinos \cite{PhysRevD.83.117702}:
In particular, Ref.~\cite{Zhou:2011rc} claimed that
the size of neutrino Yukawa couplings are severely constrained 
from the observations of supernova neutrinos 
as well as CMB radiations,
if the neutrinophilic Higgs scalar is en extremely light.
This constraint does not apply here
since the corresponding scalar is heavy enough.

%%%%%%%%%%%%%%%%%%%%%%%%%%%%%%%%%%%%%%%%%%%%%%%%%%%%%%%%%%%%%%%%%%%%%%
\subsection{Proton Decay in  Flipped $U(1)_R$ Model}
\label{sec:protondecay}
%%%%%%%%%%%%%%%%%%%%%%%%%%%%%%%%%%%%%%%%%%%%%%%%%%%%%%%%%%%%%%%%%%%%%%

Originally,
a nice feature of $U(1)_R$ symmetry is that
it explains proton stability naturally:
the dangerous baryon and lepton number violating operators
are forbidden
if we  assign $R$-charge $+1$ to all the quark and lepton superfields
and $R$-charge $0$ to the Higgses.
This nice property is not modified
when we introduce the $R$-partner Higgses $X_{0,u,d}$ with $R=2$
and the $R$-partner gauginos with $R=0$.
In our assignment, however,
we flip the $R$-charge of the right-handed neutrinos $N^c$
to be $-1$,
which may spoil the proton stability.

An operator analysis presented in Appendix~\ref{sec:operator} shows that
there is a unique dimension-five operator
that can lead to the proton decay,
\begin{align}
W_{5R} 
%\ =\ y_{Dij} Q_i D^c_j {H_d} + y_{\nu ij} N^c_i X_u L_{j} 
\ ={}-\frac{C_5^{ijkl}}{\LambdaC} U^c_i D^c_j D^c_k N^c_\ell
      \ ,
\label{eq:dim.5}
\end{align}
where 
$\LambdaC$ is a cutoff scale
and $C_5^{ijk\ell}$ is a dimensionless coefficient.
In this subsection we will give a rough estimate of the proton lifetime.
%In our calculation,
To simplify the expressions,
we will work in a flavor basis in which
the Yukawa couplings of up-type quarks and charged leptons
are diagonal;
we also define $C_5^{ijk\ell}$ to be the coefficient
of the $\ell$-th mass eigenstate of neutrinos.

%%%%%%%%%%%%%%%%%%%%%%%%%%%%%%%%%%%%%%%%%%%%%%%%%%%%%%%%%%%%%%%%%%
\begin{figure}[t]
\begin{minipage}{0.45\textwidth}
\includegraphics[width=0.95\textwidth]{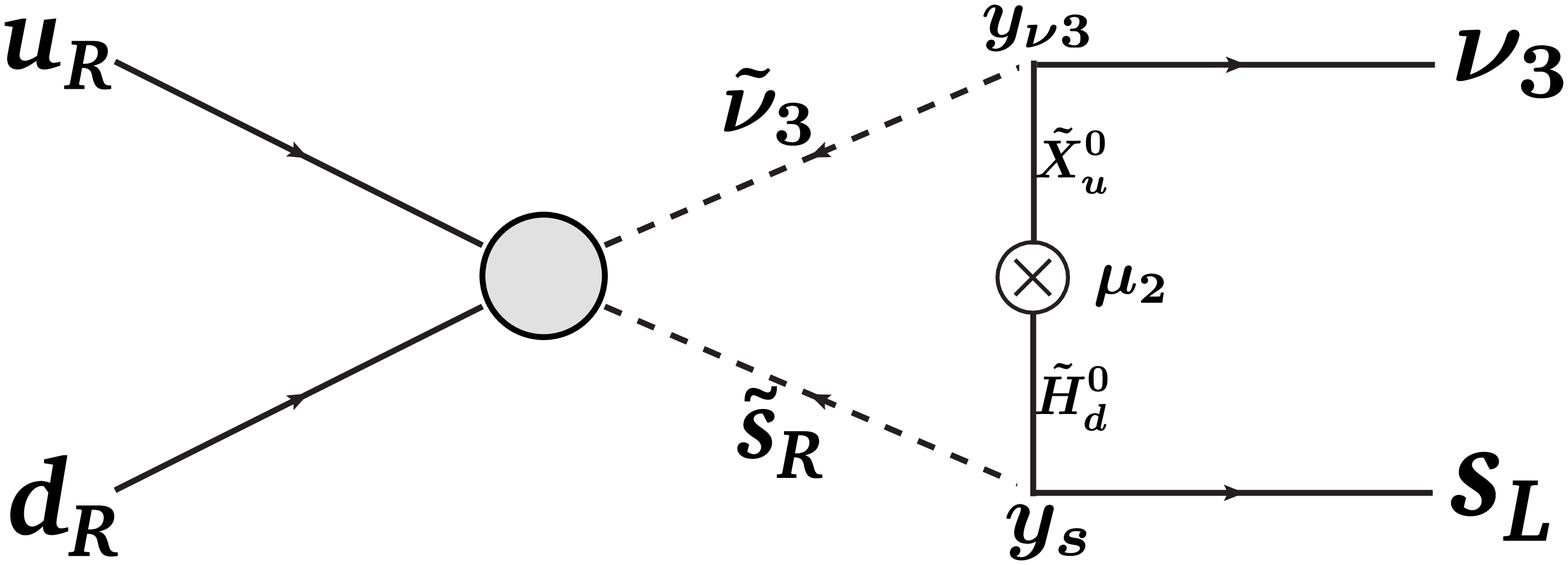}
\end{minipage}
\hspace*{0.03\textwidth}
\begin{minipage}{0.45\textwidth}
\includegraphics[width=0.95\textwidth]{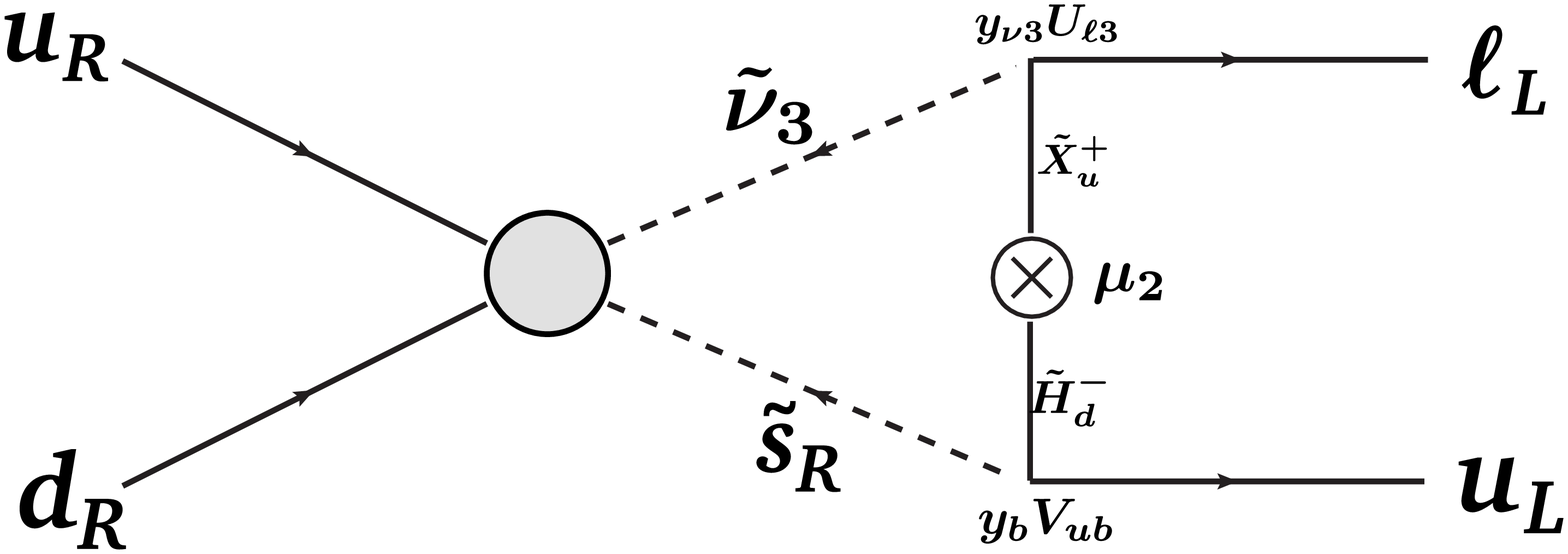}
\end{minipage}
\caption{
Left: A diagram with neutral higgsino dressing 
gives the dominant mode $p \rightarrow \bar \nu + K^+$.
Right:
Charged higgsino dressing diagrams give sub-dominant modes
$p \rightarrow \ell^+ + \pi^0$ ($\ell=e,\mu$).
For the latter, the diagrams with a sbottom $\wt{b}_R$ in the loop 
also give a comparable contribution.
}
\label{fig:diagram}
\end{figure}
%%%%%%%%%%%%%%%%%%%%%%%%%%%%%%%%%%%%%%%%%%%%%%%%%%%%%%%%%%%%%%%%%%%%%%

With the dimension five operator \eq{eq:dim.5},
the proton decay occurs
via the processes depicted in Fig.~\ref{fig:diagram},
the one with neutral Higgsino dressing 
and the other with charged Higgsino dressing.
Remarkably,
diagrams involving the top Yukawa coupling
are absent because the $\mu{}H_u H_d$ term 
is absent in the $U(1)_R$-symmetric limit,
or is extremely suppressed 
as is proportional to the tiny VEV \eq{eq:vev-x0}.
Instead, we have potentially large contributions
involving a large neutrino Yukawa coupling $y_\nu$ of
the heaviest neutrino $\nu_3$.

The partial rate
for a proton decaying into a meson $M$ and a lepton $\ell$
takes the form\footnote{
For long distance effects of RRRR-type operator, 
see Ref.~\cite{Goto:1998qg}. % for instance.
}
\begin{align}
\Gamma\!\left(p\rightarrow M + \ell \right)
\ =\ \frac{m_p}{32\pi}
     \left(1-\frac{m^2_M}{m^2_p}\right)^2 
     \frac{\abs{\alpha_p}^2}{f^2_M} 
     \,\leftB|
         \amp\!\left(p \rightarrow M  \ell \right)
       \rightB|^2
     \ .
\end{align}
Here $M_M$ and $f_M$ are the mass and the decay constant 
of the meson $M$, respectively.
The dimensionful constant $\alpha_p$ is defined through 
$\langle 0| \epsilon_{abc}(d^a_Ru^b_R)u^c_L | 0 \rangle=\alpha_p N_L $ 
where $a$, $b$, $c$ are color indices
and $N_L$ is the wavefunction of left-handed proton. 
We use the value 
$\alpha_p ={}-0.015\nunits{GeV}^3$ \cite{Aoki:1999tw}
in our calculation.
The amplitude $\amp$ 
corresponding to each diagram in Fig.~\ref{fig:diagram}
is given respectively by
\begin{align}
\amp\!\left(p \rightarrow K^+ \bar\nu_3 \right)
 &=\ 
     \frac{C_5^{1123}}{\LambdaC}\,
     \frac{y_{\nu 3} y_s \mu_2}{16\pi^2 m^2_\text{soft}}
     \ , 
\label{amp:neutralino}\\
\amp\!\left(p \rightarrow \pi^0 \ell^+\right)
 &=\ 
     \sum_{k=s,b}
     \frac{C_5^{11k3}}{\LambdaC}\,
     \frac{U_{\ell{3}}y_{\nu 3} V_{uk}y_k \mu_2}{16\pi^2 m^2_\text{soft}}
     \ , 
\label{amp:chargino}
\end{align}
where $U_{\ell 3}$ and $V_{uk}$ are mixing matrix elements,
and $m_\text{soft}$ is 
a typical mass scale of the particles propagating in the loop. 
We see that
the latter diagrams with charged higgsino dressing
are suppressed by the CKM matrix element $\cabbibo=V_{us}=0.22$,
because
$y_s\sim y_b\cabbibo^2$ and  $y_b V_{ub}\sim y_s V_{us}\sim y_b\cabbibo^3$.
Therefore the dominant decay mode is 
%that into a kaon and a neutrino 
$p\rightarrow K^+ + \bar{\nu}$.
Using the simplified expression of the neutrino Yukawa\footnote{
Here we have neglected the mild $\tan\beta$ dependence
in the denominator in \Eq{numass} to simplify the expression.
We used \Eq{numass} in our numerical calculation.
}
\begin{align}
y_{\nu}
\ \approx\ 
     \frac{m_{\nu}}{m_{3/2}}
     \left(\frac{m^2_{X_u} + \mu_2^2}{-\mu_2 v\cos\beta}\right)
     \ ,
\label{eq:Ynu:simplify}
\end{align}
we find 
the partial decay width into a kaon and an anti-neutrino is given by
\begin{align}
\Gamma\!\left(p\rightarrow K^+\bar\nu\right)
 &\approx
     \frac{m_p}{32\pi}
     \left(1-\frac{m^2_{K^+}}{m^2_p}\right)^2\!
     \frac{\alpha_p^2}{f^2_K}
     \left(\frac{C_5}{16\pi^2m_{3/2}\LambdaC}\right)^2\!\!
     \frac{m_s^2 m_\nu^2}{v^4\cos^4\beta}
     \left(\frac{\mu^2_2 + m^2_{X_u}}{m^2_\text{soft}}\right)^2
     \! ,
\end{align}
where we denote $C_5^{1123}$ simply by $C_5$.

Numerically the lifetime of proton can be estimated as
\begin{align}
\tau_p 
 = \frac{5.65\times10^{28}\nunits{yr}}{C_5^2}
    \left(\frac{\LambdaC}{2.4 \times 10^{18}\nunits{GeV}}\right)^2
    \left(\frac{m_{3/2}}{10 \nunits{eV}}\right)^2
    \left(\frac{\cos^2\beta}{0.1}\right)^2
 \ ,
\end{align}
where we took the following values of parameters:
\begin{align}
\begin{split}
m_\text{soft} &= 1.5 \nunits{TeV} \ , \qquad
\mu_2 = %300 \nunits{GeV} \ , \quad
m_{X_u} = 300 \nunits{GeV} \ ,  \qquad
\tan\beta =3 \ , %\quad
\\
m_p   &= 1.0 \nunits{GeV} \ , \qquad
\alpha_p ={}-0.015 \nunits{GeV}^3\ , 
\\
m_{K^+} &= 0.5 \nunits{GeV} \ , \qquad
f_K = 0.13 \nunits{GeV} \ , 
\\
%y_{D22}=10^{-3} \ ,
m_s   &= 0.1 \nunits{GeV} \ , \qquad
m_\nu = 0.1 \nunits{eV}  \ .
\end{split}
\label{eq:parameterset}
\end{align}
The present lower bound on the proton lifetime is 
$2.3\times10^{33}~\mathrm{yr}~\mathrm{(90\%CL)}$ obtained 
%from Super-Kamiokande experiment 
for $p \rightarrow K^+ +\bar \nu$ mode \cite{PhysRevD.72.052007}. 
By comparing the bound with the result, 
we obtain the constraint\footnote{
This applies only for $\LambdaC \agt 10^{13}\nunits{GeV}$;
otherwise, 
the operators \eq{eq:weinbergop} can not be neglected anymore.
}
%\begin{align}
%\frac{\LambdaC}{C_5} 
%\ \agt\ \frac{9.33 \times 10^{12}}{m_{3/2}} \ \mathrm{GeV} \ ,
%\end{align}
%or equivalently,
\begin{align}
\abs{C_5}
 \alt\ 
    4.96 \times 10^{-3}
    \left(\frac{\LambdaC}{2.4 \times 10^{18}\nunits{GeV}}\right)
    \left(\frac{m_{3/2}}{10 \nunits{eV}} \right)
    \left(\frac{\cos^2\beta}{0.1}\right)
    \ .
\label{eq:constraint}
\end{align}

%%%%%%%%%%%%%%%%%%%%%%%%%%%%%%%%%%%%%%%%%%%%%%%%%%%%%%%%%%%%%%%%%%%%%%
\begin{figure}[tb]
\begin{minipage}{0.95\linewidth}
\includegraphics[width=0.45\linewidth]{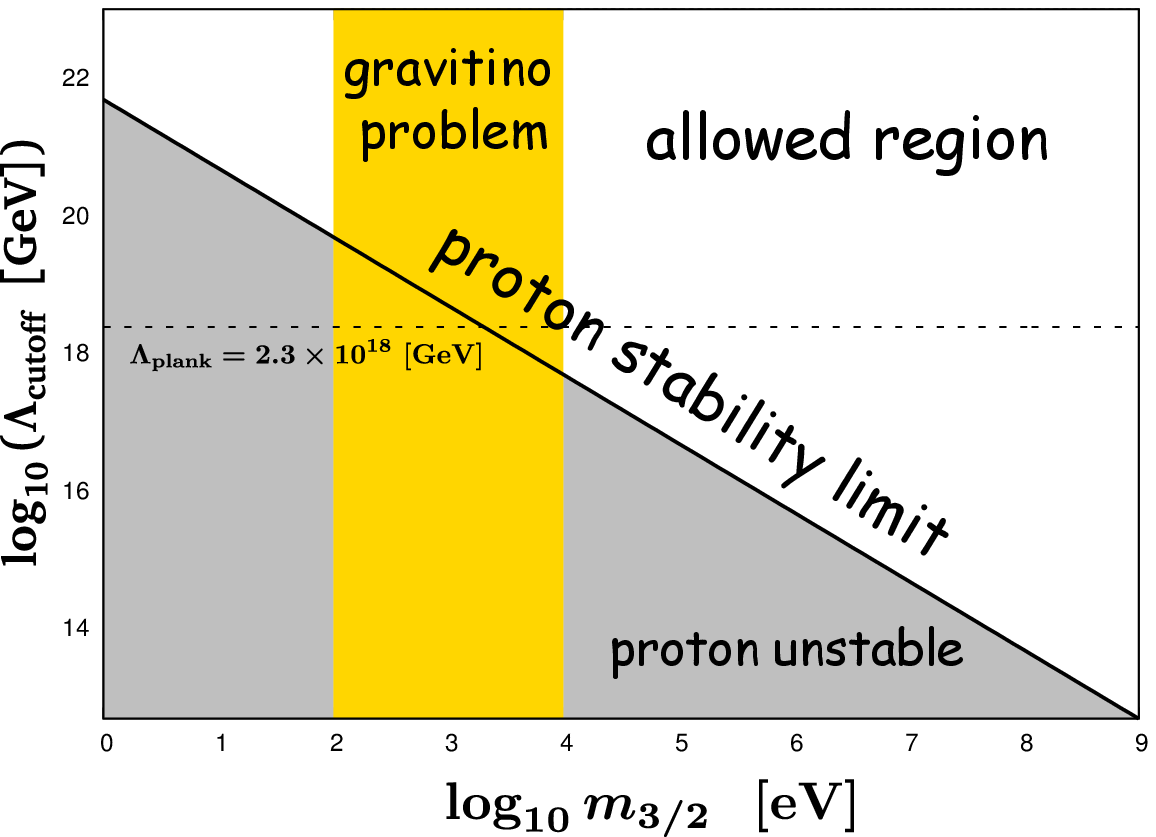} 
 \hfill
\includegraphics[width=0.45\linewidth]{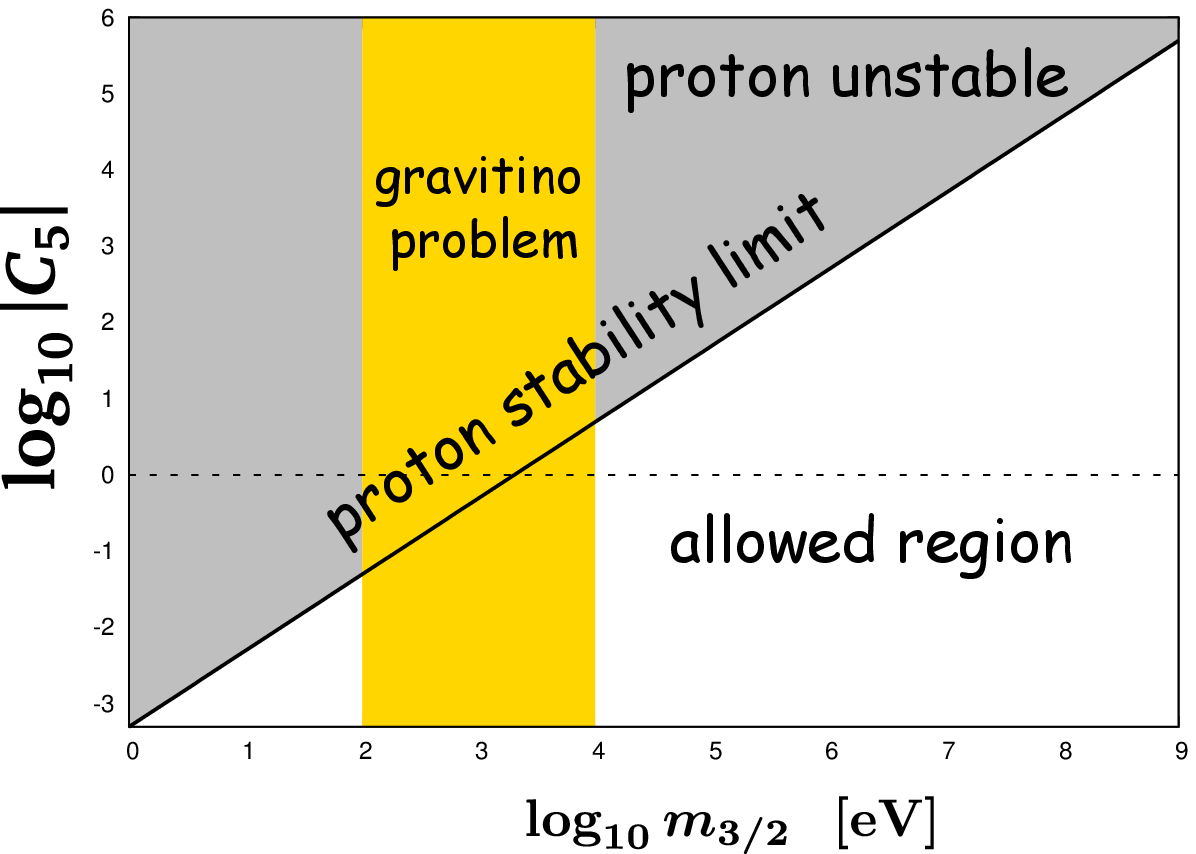} 
\end{minipage}
\caption{
Constraints from proton stability
as a function of the gravitino mass.
Left: the constraint on $\log_{10}\LambdaC$
for the fixed coupling $C_5=1$.
Right: the constraint on $\log_{10}\abs{C_5}$
for $\LambdaC=\mplanck$.
We took $\tan\beta = 3$,
$\mu_2 = 300 \nunits{GeV}$, $m_{X_u}=300\nunits{GeV}$, 
and $m_{\tilde q}= 1.5 \nunits{TeV}$.
The figures correspond to the neutrino mass 
$m_\nu = 0.1 \nunits{eV}$.
%The parameter set is given in \Eq{eq:parameterset};
%in particular, $m_\mu=0.1\nunits{eV}$ and $m_{\wt{q}}=1\nunits{T.5TeV}$.
The gray regions are excluded by the proton decay
while the yellow bands are disfavored cosmologically.
%The region below the solid line is excluded from proton decay. 
}
\label{fig:protonstability}
\end{figure}
%%%%%%%%%%%%%%%%%%%%%%%%%%%%%%%%%%%%%%%%%%%%%%%%%%%%%%%%%%%%%%%%%%%%%%

Fig.~\ref{fig:protonstability} shows
the constraint from proton stability
as a function of the gravitino mass.
The constraint is most severe for a small gravitino mass:
If the gravitino is as light as $\order{\mathrm{eV}}$,
the coefficient $C_5$ need to be a few order of magnitudes smaller than $1$ 
even when the cutoff $\LambdaC$ is equal 
to the reduced Planck scale $\mplanck$.
The constraint becomes milder for a larger gravitino mass.
This can be understood from \Eq{numass}:
For a fixed neutrino mass,
a larger gravitino mass implies a smaller neutrino Yukawa coupling,
suppressing the proton decay.
We thus find that 
the constraint can be satisfied for the gravitino mass 
larger than $10~\mathrm{keV}$,
%as large as $10~\mathrm{keV} \alt m_{3/2} \alt 10~\mathrm{MeV}$,
which is also viable cosmologically as in seen in \Eq{eq:gravitino:range}.

Notice, however, that the constraint becomes severe
for a large $\tan\beta $:
The lifetime is proportional to $\cos^4\beta$
up to a possible mild dependence in the denominator
in \Eq{numass}.
This is because 
the $X_u^0$ tadpole is proportional to the $\VEV{H_d^0}$
and because the decay amplitude involves a down-type Yukawa coupling.
Consequently,
for $\tan\beta=10$, 
%and 
%for the remaining parameters as same as in \Eq{eq:parameterset},
the proton stability 
against the dimension five operator \eq{eq:dim.5} 
requires that
the gravitino mass should be larger than $\order{100\nunits{keV}}$.
In this case,
%which corresponds to the case in which
the neutrino Yukawa coupling $y_\nu$ is
comparable to the electron Yukawa.

%**********************************************************************
%\begin{description}
%\item[$\tan\beta=3$ ($\cos^2\beta=1/10$) case]
%The constraint is rather mild.
%The gravitino mass as low as $\order{10\nunits{keV}}$
%is allowed, where $y_\nu$ is comparable to the muon Yukawa.
%\item[$\tan\beta=10$ ($\cos^2\beta=1/100$) case]
%The constraint on $\abs{C_5}$ is 10 times severer.
%The gravitino mass should be as large as $\order{100\nunits{keV}}$, 
%where $y_\nu$ is comparable to the electron Yukawa.
%\end{description}
%**********************************************************************

The above results should be compared with the MSSM case
in which the proton decay is induced by the dimension five operators
\begin{align}
\Delta{W}_{\mathrm{MSSM}}
\ ={}-\frac{C_{L}}{2\LambdaC}\,QQQL
     -\frac{C_{R}}{\LambdaC}\,U^c U^c D^c E^c 
      \ .
\label{eq:dim5:MSSM}
\end{align}
In the MSSM, the dominant contribution comes from the LLLL operator, 
and the coefficient should be 
very suppressed, $\abs{C_{L}} \alt 10^{-8}$ for $\LambdaC=\mplanck$.
Moreover, 
the RRRR operator should also be suppressed
because it can involve the top Yukawa coupling
when dressed with a charged Higgsino loop.
In our case,
the LLLL operator is absent due to the $R$-symmetry,
and with our RRRR operator,
chargino loop diagrams in Fig.~\ref{fig:diagram} 
involve a down-type Yukawa coupling,
$y_s$ or $y_b{}V_{cb}$, instead of top Yukawa.
As a consequence, the RRRR contribution in the flipped $U(1)_R$ model 
is a couple of order smaller than that in the MSSM.

%%%%%%%%%%%%%%%%%%%%%%%%%%%%%%%%%%%%%%%%%%%%%%%%%%%%%%%%%%%%%%%%%%%%%%
%\subsection{Proton Decay and Cosmological Constraints Combined}
%%%%%%%%%%%%%%%%%%%%%%%%%%%%%%%%%%%%%%%%%%%%%%%%%%%%%%%%%%%%%%%%%%%%%%

In this way, 
our flipped $U(1)_R$ assignment
can explain the smallness of neutrino masses
with relatively large Yukawa couplings
and without spoiling the proton stability so much.
The latter is certainly true for small $\tan\beta$:
for larger $\tan\beta$, on the other hand,
there is a tension between 
our mechanism for neutrino masses and the proton stability.

We note that such tension can be relaxed
if we combine the idea of flavor symmetries with the present model.
In Appendix~\ref{sec:FN},
we give an illustrative example of 
$U(1)$ flavor symmetry,
along the line if Ref.~\cite{Kakizaki:2002hs},
and show that a proper $U(1)$ charge assignment
for generating the Yukawa hierarchy
can guarantee the proton stability as well.

%%%%%%%%%%%%%%%%%%%%%%%%%%%%%%%%%%%%%%%%%%%%%%%%%%%%%%%%%%%%%%%%%%%%%%
%%%%%%%%%%%%%%%%%%%%%%%%%%%%%%%%%%%%%%%%%%%%%%%%%%%%%%%%%%%%%%%%%%%%%%
\section{Mass and Mixing of Pseudo Goldstino}
\label{sec:goldstino}
%%%%%%%%%%%%%%%%%%%%%%%%%%%%%%%%%%%%%%%%%%%%%%%%%%%%%%%%%%%%%%%%%%%%%%
%%%%%%%%%%%%%%%%%%%%%%%%%%%%%%%%%%%%%%%%%%%%%%%%%%%%%%%%%%%%%%%%%%%%%%

We now turn to another effect of $R$-symmetry breaking
in our $U(1)_R$-symmetric model.
When supersymmetry is broken in two independent sectors,
there appear two Goldstine fermions \cite{Cheung:2010mc}.
After coupling to supergravity,
one linear combination of these goldstinos becomes
the longitudinal components of the massive gravitino, 
and the other is a pseudo goldstino state, which we denote by $\zeta$.
%as a physical state in visible sector. 
%which in our case couples to the visible sector fields. 
The mass of such pseudo goldstino has been studied in literature 
\cite{Cheung:2010mc,Bertolini:2011tw,Argurio:2011hs}.

In our case, supersymmetry is broken 
in the visible sector as  well as in a hidden sector,
%As for the hidden sector SUSY breaking,
For our purpose,
we do not need to specify its precise form of the hidden sector,
but we just assume that the supersymmetry breaking is hierarchical;
the SUSY breaking scale in the hidden sector is much larger than 
that in the visible sector.
We then expect that
the physical pseudo goldstino state $\zeta$ 
resides dominantly in the visible sector.
In the limit $\lambda\rightarrow0$,
in which the EWSB is switched off, % as is seen from Eq.~\eqref{eq:EWSB}, 
the would-be goldstino in the visible sector is 
the singlino $\wt{X}_0$, the fermionic component of the singlet $X_0$,
since the visible SUSY breaking is triggered by its linear term 
in the superpotential (\ref{eq:visibleW}).
Therefore we first discuss the mass of $X_0$ fermion,
and then calculate the full neutralino mass matrix
and its smallest eigenvalue and the eigenvector.

%%%%%%%%%%%%%%%%%%%%%%%%%%%%%%%%%%%%%%%%%%%%%%%%%%%%%%%%%%%%%%%%%%%%%%
\subsection{Mass of %$X_0$ 
Singlet Fermion from $U(1)_R$ Breaking} 
\label{sec:singlino}
%%%%%%%%%%%%%%%%%%%%%%%%%%%%%%%%%%%%%%%%%%%%%%%%%%%%%%%%%%%%%%%%%%%%%%

The mass term of the singlet fermion $\wt{X}_0$ can be generated
from the contact term in the K\"ahler potential of $X_0$.
For definiteness, 
let us consider the following K\"ahler potential
\begin{align}
K(X_0,X_0^\dagger)
 &= X_0X_0^\dagger -\frac{1}{4\Lambda_0^2} {(X_0X_0^\dagger)}^2
    \ , 
    \qquad
%%\nonumber\\
%K_{X^\dagger X}^{-1}
%% &=&
%  =
%     \left(
%     1 - \frac{X_0X_0^\dagger}{\Lambda_0^2}     
%     \right)^{-1}
% \approx 
%     1 + \frac{X_0X_0^\dagger}{\Lambda_0^2}
%     \ ,
\label{eq:kahler}
\end{align}
where $\Lambda_0$ is the cutoff scale at which the contact term is generated.
%Once $\left<\int{d^2 \theta}X^2_0\right>$ takes VEVs, 
The relevant terms in the Lagrangian
 containing the $X_0$ supermultiplet are given by
\begin{align}
\mathcal{L}_{X_0}
 &= K_{X_0^\dagger X_0}F_{X_0}^\dagger F_{X_0}
    -m^2_{X_0}X_0^\dagger X_0
    -\lambda^2\left(\abs{H_u}^2+\abs{H_d}^2\right)
     X_0^\dagger X_0
\nonumber\\
 &{} 
   +\left\{
      F_{X_0}\frac{\partial W}{\partial X_0}
      - \frac12 m_{\tilde{X}_0} \wt{X}_0 \wt{X}_0 
      + \hc
     \right\}
%   - \frac12\left\{ m_{\tilde{X}_0} \wt{X}_0 \wt{X}_0 + \hc\right\}
   + \left\{ 2fm_{3/2}X_0 + \hc \right\}
       \ ,
\label{eq:lagrangian:x0}
\end{align}
where 
%$m_{X_0}^2$ is soft scalar mass from $R$-invariant mediation of SUSY breaking.
the (Majorana) mass of the would-be goldstino $\wt{X}_0$
is found to be
\begin{align}
%\mathcal{L} _{\tilde{X}_0\,\mathrm{mass}}
%  ={}- \frac12\left( m_{\tilde{X}_0} \wt{X}_0 \wt{X}_0 + \hc\right)
%       \ , \qquad
m_{\tilde{X}_0}
  ={} 
     -\frac{1}{\Lambda_0^2}
      \VEV{\int{d^2 \bar\theta}\,\frac{1}{2}X_0^\dagger X_0^\dagger}
  = 
     -\frac{\VEV{F_{X_0}}^\dagger}{\Lambda_0^2}
      \VEV{X_0}^\dagger
      \ .
\label{xmass}
\end{align}
Note that this expectation value can be non-vanishing
only in the presence of the $U(1)_R$ breaking,
the last term in the Lagrangian \eq{eq:lagrangian:x0}.
To compute it, %$X$ fermion mass \eqref{xmass}.
we use the equations of motion to get
\begin{align}
\VEV{F_{X_0}^\dagger}
 &={}-K_{X_0^\dagger X_0}^{-1}\frac{\partial W}{\partial X_0} 
\ \approx{}
      -\left(f-\lambda\VEV{H_u^0}\VEV{H_d^0}\right)
      \ , \qquad
\\
\VEV{X_0}
 &=  \frac{2 f}{m^2_{X_0} + \lambda^2v^2 + \delta{m_{X_0}^2}}
      \,m_{3/2}
      \ .
\end{align}
Here in the first equation, we have used the fact that
the K\"ahler metric is almost canonical
since 
%the $R$-breaking VEV of $X_0$ is very small, 
$\abs{\VEV{X_0}}\ll\Lambda_0$.
In the second equation, we note that
the previous result \eqref{eq:vev-x0} 
is slightly modified
by the soft scalar mass $\delta{m_{X_0}^2}$
due to the contact term in Eq.~\eqref{eq:kahler},
\begin{align}
\delta{m_{X_0}^2}
\ \equiv \ 
     \frac{\abs{\VEV{F_{x_0}}}^2}{\Lambda_0^2}
\ \approx\    
     \frac{1}{\Lambda_0^2}
     \abs{f-\lambda v^2\sin\beta\cos\beta}^2
     \ .
\end{align}
By plugging the VEVs of $X_0$ and $F_{X_0}$ into the expression \eq{xmass},
we thus find that the $X_0$ fermion mass is proportional to the gravitino mass
as
\begin{align}
m_{\tilde{X}_0}
\ \approx\ 2\suppression_1m_{3/2}  \ , \qquad
\suppression_1 
\ \equiv \ 
    \frac{f\left(f-\lambda v^2 \sin\beta\cos\beta\right)
          }{\Lambda_0^2
            \left(m^2_{X_0} + \lambda^2v^2 + \delta{m_{X_0}^2}\right)
          }
    \ .
\label{eq:gamma1}
\end{align}

The result \eqref{eq:gamma1} is consistent with
the general assertion \cite{Cheung:2010mc} that
the pseudo goldstino mass is twice the gravitino mass
in the ``sequestered'' limit.
This can be seen as follows:
If the soft mass of $X_0$ is zero, $m^0_{X_0} \rightarrow 0$,
and if the Higgs sector decouples from the visible SUSY breaking,
$\lambda \rightarrow 0$,
we have $\delta{}m_{X_0}^2\rightarrow{}f^2/\Lambda_0^2$, and hence,
\begin{align}
\suppression_1 \longrightarrow 1 \ .
\end{align}
%in accordance with the general argument \cite{Cheung:2010mc}.
Otherwise,
the pseudo goldstino mass is suppressed by a factor $\suppression_1$,
which is roughly of order $f/\Lambda_0^2$,
in accordance with the general argument \cite{Bertolini:2011tw}.
For instance, 
for a moderate choice of parameters
$f \sim \left(10^3\nunits{GeV}\right)^2$ and
$\Lambda_0 \sim 10^7\nunits{GeV}$,
the mass of the pseudo goldstino is smaller than 
$\order{10\nunits{eV}}$ 
even if the gravitino is as heavy as $m_{3/2} \sim \mathrm{GeV}$.

%%%%%%%%%%%%%%%%%%%%%%%%%%%%%%%%%%%%%%%%%%%%%%%%%%%%%%%%%%%%%%%%%%%%%%
\subsection{Neutralino Mass Matrix with Dirac Gaugino}
%%%%%%%%%%%%%%%%%%%%%%%%%%%%%%%%%%%%%%%%%%%%%%%%%%%%%%%%%%%%%%%%%%%%%%

Now we consider the neutralino mass matrix 
which incorporates the pseudo goldstino.
In this subsection,
let us denote by $\chi_B$ and $\chi_W$ 
the Dirac partner of the bino $\wt{B}$ and the wino $\wt{W}$,
respectively.
The Dirac mass terms for $SU2)\times U(1)$ gauginos are
\begin{align}
\mathcal{L}_{\mathrm{gaugino}} 
\ ={}-m_{\wt{B}} \chi_B \wt{B}- m_{\wt{W}} \chi_W \wt{W}
     + \hc \ .
\end{align}
%where the Dirac masses for bino and wino are denoted by $m_\wt{B}$
Then, in a basis given by
%We write the neutralino mass matrix $\mathcal{M}_{\wt{N}}$
%in a basis given by
\begin{eqnarray}
\vec{\Psi}^T
% &=&  \left(
%       \chi_B,\chi_{W^3},\, 
%       \wt{W}^3\,\wt{B}, ;\,
%       \wt{H}_d^0, \wt{H}_u^0, \,
%       \wt{X}_d^0, \wt{X}_u^0\,;\,
%       \wt{X}_0 
%       \right)
%\non\\
 &=&  \left(\begin{array}{cccc|cccc|c}
        \makebox[23pt]{$\chi_B$}
      & \makebox[23pt]{$\chi_{W^3}$}
      & \makebox[25pt]{$\wt{W}^3$}
      & \makebox[25pt]{$\wt{B}$}
      & \makebox[26pt]{$\wt{H}_d^0$}
      & \makebox[26pt]{$\wt{H}_u^0$}
      & \makebox[26pt]{$\wt{X}_d^0$}
      & \makebox[26pt]{$\wt{X}_u^0$}
      & \makebox[26pt]{$\wt{X}_0$}
       \end{array}\right)
     \ ,
\label{eq:basis}
\end{eqnarray}
the neutralino mass matrix takes the form
\begin{align}
\mathcal{M}_{\tilde{N}} &= 
\left(
\begin{array}{cc|cc|cc|cc|c}
 0      & 0      & 0      & \mbino & 0 & 0 & 0 & 0 & 0 \\ 
 0      & 0      & \mwino & 0      & 0 & 0 & 0 & 0 & 0 \\
\hline
 0      & \mwino & 0      & 0      & a m_Z & b m_Z & a'm_V & b'm_V & 0 \\
 \mbino & 0      & 0      & 0      & c m_Z & d m_Z & c'm_V & d'm_V & 0 \\
\hline
 0 & 0 & a m_Z & c m_Z & 0 & -\lambda v_0 & 0 & +\mu_2 & -\lambda v_u \\
 0 & 0 & b m_Z & d m_Z & -\lambda v_0 & 0 & -\mu_1 & 0 & -\lambda v_d \\
\hline
 0 & 0 & a'm_V & c'm_V & 0 & -\mu_1 & 0 & 0 & 0 \\
 0 & 0 & b'm_V & d'm_V & +\mu_2 & 0 & 0 & 0 & 0 \\
\hline
 0 & 0 & 0 & 0 & -\lambda v_u & -\lambda v_d & 0 & 0 & \mxino
 \end{array}\right) \ ,
%\label{eq:R-mass}
\label{eq:massmatrix:full}
\end{align}
where
$\mxino=2m_{3/2}\suppression_1$,
and we have defined $\VEV{X^0}=v_0$ and 
\ba
m^2_V
 \equiv \frac{g_1^2+g_2^2}{2}
        \left(\abs{\VEV{X^0_d}}^2+\abs{\VEV{X^0_u}}^2\right)
        \ , \qquad
\tan\gamma
 \equiv \frac{\VEV{X^0_u}}{\VEV{X^0_d}}
        \ .
\ea
We have also used the following abbreviations:
with $\theta_W$ being the weak mixing angle,
\begin{align}
U\ \equiv\ 
   \begin{pmatrix}
     a & b \\
     c & d
   \end{pmatrix}
 &\equiv
      \left(\begin{array}{rr}
          c_\beta c_W & -s_\beta c_W \\
         -c_\beta s_W &  s_\beta s_W 
      \end{array}\right)
 =    \left(\begin{array}{r}
          \cos\theta_W \\
         -\sin\theta_W 
      \end{array}\right)
      \left(\begin{array}{rr}
          \cos\beta & -\sin\beta
      \end{array}\right) \ ,
\label{eq:Udef}
%\\
%U'\ \equiv\ 
%   \begin{pmatrix}
%     a' & b' \\
%     c' & d'
%   \end{pmatrix}
% &\equiv
%\left(\begin{array}{rr}
%   c_\gamma c_W & -s_\gamma c_W \\
% - c_\gamma s_W &  s_\gamma s_W 
%\end{array}\right)
% =    \left(\begin{array}{r}
%          \cos\theta_W \\
%         -\sin\theta_W 
%      \end{array}\right)
%      \left(\begin{array}{rr}
%          \cos\gamma & -\sin\gamma
%      \end{array}\right) \ ,
%\label{eq:Ugamma:def}
\end{align}
where $\VEV{H^0_d}=v\cos\beta$ and $\VEV{H^0_u}=v\sin\beta$:
the primed quantities are obtained
by replacing $\beta$ with the angle $\gamma$ defined above.

The mass eigenvalue $\Omega_D$ can be found
by solving the characteristic equation 
\begin{align}
0\ =\ \mathrm{det} 
      \left(\mathcal{M}_{\tilde{N}} - \Omega_D \hat{1}\right)
 \ =\ \mathrm{det} \mathcal{M}_{\tilde{N}} 
     -\Omega_D\,\mathrm{det}\mathcal{M}_{\tilde{N}} 
      \mathrm{Tr}\!\left[\mathcal{M}_{\tilde{N}}^{-1}\right] 
     + \cdots \ , 
\label{eq:characteristic}
\end{align}
%%%
where the dots represents terms higher order in $\Omega_D$.
The first and the second terms are calculated to be
%%%
\begin{align}
\mathrm{det}\mathcal{M}_{\tilde{N}}
 &= 2\suppression_1 m_{3/2}\mu^2_1 \mu^2_2 m^2_{\wt{B}} m^2_{\wt{W}}
    \ , 
\\
\mathrm{det}\mathcal{M}_{\tilde{N}} 
\mathrm{Tr} [\mathcal{M}_{\tilde{N}}^{-1}]
 &= m^2_{\wt{B}} m^2_{\wt{W}}
    \left[
      \mu^2_1\mu^2_2
      +\lambda^2v^2
       \left(\mu^2_1\sin^2\beta+\mu^2_2\cos^2\beta\right)
     \right]
     \ ,
\end{align}
%%%
up to $\mathcal{O}(m_{3/2}^2)$ and $\mathcal{O}(m_{3/2})$, respectively.
%
%the pseudo goldstino mass is proportional to the gravitino mass and 
%be small if gravitino is very light. 
%
The lowest eigenvalue,
which we identify with the pseudo goldstino mass $m_\zeta$,
can be calculated 
by keeping only the first two terms in \Eq{eq:characteristic},
%For the smallest eigenvalue, we can neglect the higher order terms in 
%\eqref{eq:characteristic} and obtain
%%%
\begin{align}
%\Omega_D 
m_\zeta
\ =\ 2m_{3/2} \suppression_1\suppression_2 
%+ \order{m_{3/2}^2}
     \ ,
\end{align}
where %\red{(Please write $\gamma_1$)}
the suppression factor
$\suppression_1$ is given in Eq.~\eqref{eq:gamma1} and
%%%
\begin{align}
%\suppression_2
%\ =\ \frac{\mu^2_1 \mu^2_2}%
%          {\mu^2_1 \mu^2_2 
%          +\lambda^2 v^2 ( \mu^2_1 \sin^2{\beta} + \mu^2_2\cos^2{\beta})
%          } \ , \quad
\suppression_2^{-1}
\ =\ 1+\frac{\lambda^2 v^2\sin^2{\beta}}{\mu_2^2}
      +\frac{\lambda^2 v^2\cos^2{\beta}}{\mu_1^2}
     \ .
\end{align}
We see that the mass of the true pseudo goldstino 
is suppressed by a factor $\suppression_1\suppression_2$.
%Since the latter factor $\suppression_2$ is of order $1$,
%the huge suppression by
%$\suppression_1\sim {f/\Lambda_0^2}$ is not modified so much.
We note that
the lightness of the pseudo goldstino
will be protected against quantum corrections
by the $U(1)_R$ symmetry in the visible sector,
as long as Dirac nature of the gauginos is kept.

The corresponding eigenvector can be found
perturbatively in $R$-breaking effects.
By writing
%%%
$
   \vec{\Psi}_{\zeta}
=  \vec{\Psi}_{\zeta}^{(0)}
  +\vec{\Psi}_{\zeta}^{(1)}
% +\order{m_{3/2}^2}
  +\cdots
$,
we find 
\begin{align}
  \vec{\Psi}_{\zeta}^{(0)T}
 &= 
     \sqrt{\suppression_2}
     \left(\begin{array}{cc|cc|cc|c|c}
        \makebox[12pt]{$0$}
      & \makebox[12pt]{$0$}
      & \makebox[12pt]{$0$}
      & \makebox[10pt]{$0$}
      & \makebox[12pt]{$0$}
      & \makebox[12pt]{$0$}
      &
        \makebox[30pt]{$\vec{\zeta}_{\wt{X}}^{(0)T}$}
%      &
      & \makebox[10pt]{$1$}
       \end{array}\right)
     \ ,
\non\\
  \vec{\Psi}_{\zeta}^{(1)T}
 &= \sqrt{\suppression_2}
    \left(\begin{array}{c|cc|c|cc|c}
        \makebox[30pt]{$\vec{\zeta}_{\chi}^{(1)T}$}
      & \makebox[12pt]{$0$}
      & \makebox[10pt]{$0$}
      &
        \makebox[30pt]{$\vec{\zeta}_{\wt{H}}^{(1)T}$}
      & \makebox[12pt]{$0$}
      & \makebox[12pt]{$0$}
      & \makebox[12pt]{$0$}
       \end{array}\right)
%\non\\
% &= \sqrt{\suppression_2}
%    \left(\begin{array}{cccc|cccc|c}
%        \makebox[20pt]{$\zeta_{\chi_B}$}
%      & \makebox[22pt]{$\zeta_{\chi_W}$}
%      & \makebox[8pt]{$0$}
%      & \makebox[10pt]{$0$}
%      & \makebox[22pt]{$\zeta_{\wt{H}_d}$}
%      & \makebox[20pt]{$\zeta_{\wt{H}_u}$}
%      & \makebox[8pt]{$0$}
%      & \makebox[12pt]{$0$}
%      & \makebox[12pt]{$0$}
%       \end{array}\right)
     \ .
\end{align}
The two component vectors
$\vec{\zeta}_{\wt{X}}^{(0)}$, 
$\vec{\zeta}_{\chi}^{(1)}$ and $\vec{\zeta}_{\wt{X}}^{(1)}$
%are calculated in App.~\ref{sec:mixing}
%and 
are given 
%by \Eq{eq:eigenvector:xino},
%\eq{eq:eigenvector:chi} and \eq{eq:eigenvector:higgsino},
respectively by
\begin{align}
\vec{\zeta}_{\wt{X}}^{(0)}
\ =\ 
    \left(\begin{array}{r}
        {}-\frac{\lambda vc_\beta}{\mu_1} \\
        {}+\frac{\lambda vs_\beta}{\mu_2}
    \end{array}\right)
    \ , \qquad
\vec{\zeta}_{\chi}^{(1)}
\ =\ 
     m_{3/2}
     \left(\begin{array}{r}
          - \frac{\sin\theta_W}{\mbino} \\
            \frac{\cos\theta_W}{\mwino}
      \end{array}\right)
      \suppression_3
      \ , \qquad
\vec{\zeta}_{\wh{H}}^{(1)}
\ =\ 
     m_\zeta 
    \left(\begin{array}{c}
         \frac{\lambda v s_\beta}{\mu_2^2} \\
         \frac{\lambda v c_\beta}{\mu_1^2} 
    \end{array}\right) 
     \ ,
\label{eq:goldstino:comp}
\end{align}
where the dimensionless factor $\suppression_3$ is defined by
\begin{align}
\suppression_3
 &={}
     -m_Zs_\beta c_\beta 
    \left(\frac{\lambda v}{\mu_2^2}
         -\frac{\lambda v}{\mu_1^2}
    \right)
    \frac{m_\zeta}{m_{3/2}}
     +\lambda v
    \left(
        \frac{s_\beta s_\gamma}{\mu_2}
       +\frac{c_\beta c_\gamma}{\mu_1}
    \right) 
    \frac{m_V}{m_{3/2}}
    \ .
\label{eq:goldstino:suppression}
\end{align}
We see that 
at the leading order,
only the xinos $\wt{X}_{d,u}^0$ can mix 
with the would-be goldstino $\wt{X}_0$
because of the $U(1)_R$ charge conservation;
at the next order,
the $\wt{X}_0$ mixes with the fermions with $R=-1$,
that is, the higgsinos $\wt{H}_{d,u}^0$
and the gaugino Dirac partners $\chi_{B,W}$,
since the gravitino mass $m_{3/2}$ has $R={}-2$
in the sense of spurion.

In the sequestered limit, 
we have $\suppression_1\rightarrow 1$,
$\suppression_2\rightarrow 1$ and $\suppression_3\rightarrow 0$,
so that the eigenvalue and the eigenvector reduce to 
\begin{align}
  m_\zeta
\ \longrightarrow\ 2m_{3/2}
    \ , \qquad 
  \vec{\Psi}_{\zeta}^T
\ \longrightarrow 
%      \left( 0,0,0,0\,;\,0,0,0,0\,;\,1\right)
      \left(\begin{array}{cc|cc|cc|cc|c}
           0 & 0 & 0 & 0 & 0 & 0 & 0 & 0 & 1
           \end{array}
      \right)
      \ ,
\end{align}
as is expected.
In other cases,
the mass eigenvalue $m_\zeta$ is much suppressed,
and hence,
%We also observe that
the higgsino component $\vec{\zeta}_{\wt{H}}^{(1)}$ 
is more suppressed than that of
the gaugino Dirac partner, $\vec{\zeta}_{\chi}^{(1)}$:
\begin{align}
\vec{\zeta}_{\wt{H}}^{(1)}
\ =\ \order{\frac{m_\zeta}{m_{\mathrm{weak}}}}
     \ , \qquad
\vec{\zeta}_{\chi}^{(1)}
\ =\ \order{\frac{m_{3/2}}{m_{\mathrm{weak}}}}
     \ .
\end{align}

%%%%%%%%%%%%%%%%%%%%%%%%%%%%%%%%%%%%%%%%%%%%%%%%%%%%%%%%%%%%%%%%%%%%%%y
%%%%%%%%%%%%%%%%%%%%%%%%%%%%%%%%%%%%%%%%%%%%%%%%%%%%%%%%%%%%%%%%%%%%%%
\subsection{Implications on Higgs Phenomenology and Cosmology}
\label{sec:implications}
%%%%%%%%%%%%%%%%%%%%%%%%%%%%%%%%%%%%%%%%%%%%%%%%%%%%%%%%%%%%%%%%%%%%%%
%%%%%%%%%%%%%%%%%%%%%%%%%%%%%%%%%%%%%%%%%%%%%%%%%%%%%%%%%%%%%%%%%%%%%%

%Some remarks are in order here
%concerning the coupling 
%of the pseudo goldstino and its scalar partner
%to the $125\nunits{GeV}$ Higgs particle.

Let us briefly discuss implications 
%of the results obtained here
on the Higgs phenomenology and cosmology.

First,
the invisible decay width of the $125\nunits{GeV}$ Higgs 
into a pair of the pseudo goldstinos
is highly suppressed due to the approximate $R$-symmetry.
This is because
fields with $R$-charge assignment other than $+1$ contain 
a small goldstino component with suppression factor 
$m_{2/3} / m_\text{weak}$.
The Higgs decay into a pseudo goldstino and other neutralino
is kinematically forbidden if the neutralino other than
the pseudo goldstino are heavier than the lightest Higgs boson.

The second remark is concerned with the invisible cascade decay
of the $125\nunits{GeV}$ Higgs.
If the scalar partner $\phi$ of the pseudo goldstino is so light that
the decay mode of the Higgs into a pair of $\phi$
is kinematically allowed,
the cascade decay into gravitinos and pseudo goldstinos 
$h^0 \rightarrow \phi\phi \rightarrow \zeta {\tilde G} \zeta {\tilde G}$ 
dominates almost entire Higgs decay channel \cite{Bertolini:2011tw},
which contradicts with recent LHC discovery \cite{Giardino:2012dp}.
In our case,
the scalar goldstino $\phi$ receives a mass $\lambda^2 v^2$ 
from the visible sector SUSY breaking,
in addition to the hidden-sector soft mass.
Therefore,
the invisible cascade decay of the Higgs is kinematically forbidden,
if we assume $\lambda> 0.36$,
or if the scalar pseudo goldstino has a soft mass larger 
than the half the Higgs mass.

%\red{(Some facts or implications drawn 
%from the pseudo goldstino mass and the corresponding eigenvalues  
%should be written here.)}

Next we briefly discuss the cosmological constraints
on the present model.
%In the present model,
%the lightest supersymmetric particle (LSP) is the pseudo goldstino 
%whereas the next-to-LSP (NLSP) is the gravitino.
%
Since we are supposing the low-scale SUSY breaking in the visible sector,
the lightest observable-sector supersymmetric particle (LOSP) 
mainly decays into a pseudo goldstino rather than a gravitino.
This fact helps us avoid the constraint from 
the BBN \cite{Cheung:2010mc}.

We should still worry about the overproduction problem,
both for the gravitino and the pseudo goldstino.
Let us first discuss the gravitino case.
If gravitinos are 
thermally produced in the early Universe,
the estimated abundance easily exceeds the limit $\Omega{}h^2 < 0.1$
for the gravitino heavier than $\order{100\nunits{eV}}$.
On the other hand,
if gravitinos are not thermalized \cite{Endo:2007sz},
the overproduction excludes the mass smaller than $10\nunits{keV}$,
whereas the gravitino heavier than $\order{10\nunits{keV}}$ can evade
the overproduction if the reheating temperature is sufficiently low\footnote{
If the reheating temperature is low, 
the thermal leptogenesis is difficult to be achieved. 
A possibility of electroweak baryogenesis in $R$-symmetric models
as was recently discussed in Ref.~\cite{Fok:2012fb}.
}.
Moreover,
for the mass smaller than $\order{100\nunits{eV}}$,
the allowed region is much reduced to
$m_{3/2} \alt 16 \ \mathrm{eV}$
due to a warm dark matter constraint \cite{Viel:2005qj,Boyarsky:2008xj}.
%As a result,
Putting it all together,
%in order to meet the cosmological constraints,
we must assume that the gravitino mass satisfies 
either $m_{3/2} \alt 16 \ \mathrm{eV}$ 
or $10 \ \mathrm{keV} \alt m_{3/2}$, % \alt 10 \ \mathrm{MeV}$.
as was announced in \Eq{eq:gravitino:range}
% \cite{Shirai:2010rr}.

In contrast,
the pseudo goldstino couples a bit strongly to the MSSM particles 
and is expected to be thermalized.
Recalling that 
its mass $m_\zeta$ is much suppressed 
compared to the gravitino mass $m_{3/2}$ as in \Eq{eq:gamma1},
the pseudo goldstino will be lighter than $16\nunits{eV}$ 
for the gravitino mass $m_{3/2}\agt 10\nunits{keV}$;
depending on the suppression factor $\suppression_1$,
it may be possible that the pseudo goldstino evades
the overclosure problem
even when the gravitino mass is as large as $\order{10\nunits{MeV}}$.
If so,
the pseudo goldstino can give only a small contribution 
to the relic density.

In the present model,
the lightest supersymmetric particle (LSP) is the pseudo goldstino 
whereas the next-to-LSP (NLSP) is the gravitino,
which can decay into a photon and a pseudo goldstino.
In this situation,
we should also be careful about the constraints from 
the gamma-ray line search from the Galactic Center region 
and the overproduction of the isotropic diffuse photon background;
there is a stringent limit \cite{Yuksel:2007dr} 
on mono-energetic photons emitted from decaying particles.
In our case, we expect that 
such decay of the gravitino %into a photon and a pseudo goldstino
is very suppressed
since the pseudo goldstino mass eigenstate
has very small gaugino and higgsino components.
If this is really the case,
the gravitino is sufficiently long-lived
and the present model can evade the constraint
from diffuse gamma-ray line search.
A preliminary consideration shows that 
only the gravitino heavier than $\order{100\nunits{MeV}}$ 
would be excluded 
even if the mixing between the photino and the pseudo goldstino
is of order of $\order{m_{3/2}/m_{\mathrm{weak}}}$.

\section{Conclusion and Discussion}
\label{sec:conclusion}
%%%%%%%%%%%%%%%%%%%%%%%%%%%%%%%%%%%%%%%%%%%%%%%%%%%%%%%%%%%%%%%%%%%%%%
%%%%%%%%%%%%%%%%%%%%%%%%%%%%%%%%%%%%%%%%%%%%%%%%%%%%%%%%%%%%%%%%%%%%%%

We have studied the $U _R(1)$ symmetric extension of SUSY standard model
which contains Dirac gauginos and extended Higgs sector, assuming that 
the Dirac gaugino masses are induced from hidden sector SUSY breaking
while the extended Higgs sector incorporates the visible SUSY breaking.
%The basis assumption in the present work is that
The $U(1)_R$ symmetry in the visible sector
is broken solely by the minimal coupling to supergravity,
so that $U(1)_R$ breaking effects are proportional to the gravitino mass.
After the EWSB,
the $R$-charged Higgs fields develop nonzero VEVs 
due to the tadpole terms induced from the $U(1)_R$ breaking.
The VEVs are proportional to the gravitino mass 
and hence can be very small 
when the gravitino mass is much smaller than the weak scale.

We have then shown 
the generation of small neutrino masses
through the $U(1)_R$ breaking effects.
With our flipped $U(1)_R$ charge assignment,
the right-handed neutrinos couple to the $R$-partner Higgs field $X_u$
that develops the very small VEV. 
Therefore the small neutrino masses can naturally be obtained
even if the neutrino Yukawa couplings are of order unity,
provided that the gravitino mass is as low as $1$---$10\nunits{eV}$.
We have also examined how our generation mechanism
of neutrino mass is constrained from the proton decay
induced by the dimension five operator, $U^c D^c D^c N^c$, 
allowed by our $U(1)_R$ charge assignment.
Interestingly,
the amplitude for the dominant decay mode
involves the neutrino Yukawa coupling, instead of the top Yukawa.
Therefore the constraint from proton stability
is severer for a larger neutrino Yukawa coupling,
which implies a small gravitino mass for a fixed neutrino mass.
We have estimated the lifetime of protons
and found that
the gravitino should be heavier than $1\nunits{keV}$
if we require the coefficient $C_5$ of the dimension five operator
to be of order unity
and the cutoff $\LambdaC$ to be the Planck scale.
Actually, 
%this region of the gravitino mass is cosmologically constrained;
%in fact, 
the gravitino mass between $10\nunits{eV}<m_{3/2}<10\nunits{keV}$
is cosmologically excluded %
\cite{Ichikawa:2009ir, Viel:2005qj, Boyarsky:2008xj},
and hence our model for the neutrino mass generation
evade the constraint from proton decay and is cosmologically safe
if gravitino is as heavy as $10$---$100\nunits{keV}$.
We have also suggested how the constraint from the proton stability
can be relaxed if we adopt the $U(1)$ flavor symmetry
%Froggatt--Nielsen mechanism
for generating the hierarchical structure of Yukawa matrices.

Another significant effect of the $U(1)_R$ breaking
is the mass and the mixing of the pseudo goldstino,
which is Nambu-Goldstone fermion of the visible SUSY breaking
and is massless in the $U(1)_R$ symmetric limit.
We have analyzed the neutralino mass matrix that 
incorporates the Dirac mass parameters of gauginos and higgsinos,
and obtained to the lowest order in $R$-breaking
the smallest mass eigenvalue and the corresponding mixing angles 
to other neutralinos.
According to the general argument, 
the mass of the pseudo goldstino in the sequestered limit 
is twice the gravitino mass.
In our case, 
the pseudo goldstino has a coupling to the Higgs fields,
and consequently, 
its mass is proportional to but suppressed from the gravitino mass.
In our concrete calculation,
in which the pseudo goldstino gets a mass
from the contact term in the K\"ahler potential,
the suppression factor is very small
if the cutoff scale $\Lambda_0$ of the contact term 
is much larger than the weak scale.
%We expect that
%(in the case of Dirac gauginos,)
%the lightness of the pseudo goldstino
%will be protected against quantum corrections
%by the $U(1)_R$ symmetry in the visible sector.
%\red{
For the mixings, we have found that
the pseudo goldstino have suppressed contaminations
of gauginos and higgsinos.
%}

%On the other hand,
As we mentioned above,
the gravitino as heavy as $100\nunits{keV}$ 
is cosmologically safe
and also satisfies the proton stability.
In this case,
the gravitino is the NLSP while the LSP is the pseudo goldstino,
which can be as light as $10\nunits{eV}$, 
or even much lighter. % than $1\nunits{eV}$,
%depending on the suppression factor.
It is then tempting to speculate that
the gravitino could be a part of cold dark matter,
%which decays into the pseudo goldstino,
and that the pseudo goldstino could play a role of 
%a warm dark matter, or 
dark radiation.
To examine such possibilities, % more carefully,,
%it will be important to investigate
we should investigate
the properties of the pseudo goldstino more extensively,
for instance,
by including the anomaly mediated gaugino masses
and other quantum corrections to the mass matrix.
This will be important 
also for a quantitative study of diffuse gamma-ray line.

%\red{
%************************************************************
%}
%
%One should be careful, however,
%about another constraint coming from
%diffused gamma-ray line search.
%since a gravitino can decay into a photon and a pseudo goldstino.
%
%
%However, if gravitino mass scale exceeds $10\nunits{MeV}$,
%the model is excluded by diffused gamma-ray line search.
%
%
%
%To examine such possibility,
%it will be important to examine
%whether the cosmological constraints from diffused gamma-ray line search.
%
%
%As a step,
%we have examined to some details
%the mass and the mixing of the pseudo goldstino
%
%
%
%
%
%\red{
%************************************************************
%}

Although we have focused in the present work
on the constraints from the proton stability
and also on some cosmological ones,
there remain many issues to be discussed:
%a partial list includes
the constraints from EW precision measurements
and an implementation of the $125\nunits{GeV}$ Higgs 
should be examined more intensively:
phenomenological impacts of large neutrino Yukawa couplings
as in other ``neutrinophilic'' models
may deserve further study
in our $U(1)_R$ symmetric model.
%\red{
%In addition,
%it may be important to investigate
%the properties of the pseudo goldstino
%by including the anomaly mediated gaugino masses.
%}
The assumptions behind our present setup
should also be elucidated:
a partial list includes
the origin of the hidden sector SUSY breaking,
the UV completion of the visible SUSY breaking sector
and the embedding into a semi-simple gauge group.
In particular,
the smallness of the gravitino mass
should be examined in a consistent framework,
although we have treated it simply as a free parameter.
In this respect, 
it may be possible that
the smallness of gravitino mass
as well as the absence of certain operators 
can be explained
in the (warped) extra-dimensional setup 
along the line of Ref.~\cite{Chacko:2004mi}.

%%%%%%%%%%%%%%%%%%%%%%%%%%%%%%%%%%%%%%%%%%%%%%%%%%%%%%%%%%%%%%%%%%%%%%
%\end{document}
%%%%%%%%%%%%%%%%%%%%%%%%%%%%%%%%%%%%%%%%%%%%%%%%%%%%%%%%%%%%%%%%%%%%%%

%%%%%%%%%%%%%%%%%%%%%%%%%%%%%%%%%%%%%%%%%%%%%%%%%%%%%%%%%%%%%%%%%%%%%%
\section*{Acknowledgement}
\noindent
The authors are grateful
to Shigeki~Matsumoto, Takehiko~Asaka, and Hiroaki~Nagao 
for useful comments and discussions on cosmological constraints on gravitino.
They would like to thank Yuichiro~Nakai 
for useful discussion on visible SUSY breaking
and Yusuke~Shimizu 
for valuable discussions about flavor physics.
This work was supported in part by the Grant-in-Aid for the Ministry of 
Education, Culture, Sports, Science, and Technology, Government of Japan 
(No.~23740190 [T.S.] and No.~22540269 [H.N.]).
This work was also supported by 
Strategic Young Researcher Overseas Visits Program 
for Accelerating Brain Circulation in JSPS (T.S.).

%\newpage
%%%%%%%%%%%%%%%%%%%%%%%%%%%%%%%%%%%%%%%%%%%%%%%%%%%%%%%%%%%%%%%%%%%%%%
%%%%%%%%%%%%%%%%%%%%%%%%%%%%%%%%%%%%%%%%%%%%%%%%%%%%%%%%%%%%%%%%%%%%%%
\appendix

%%%%%%%%%%%%%%%%%%%%%%%%%%%%%%%%%%%%%%%%%%%%%%%%%%%%%%%%%%%%%%%%%%%%%%
\section{Baryon and Lepton Number Violating Operators}
\label{sec:operator}
%%%%%%%%%%%%%%%%%%%%%%%%%%%%%%%%%%%%%%%%%%%%%%%%%%%%%%%%%%%%%%%%%%%%%%

In this appendix,
we systematically study baryon and/or lepton number violating operators
allowed by the gauge and $U(1)_R$ symmetries of the present model.

We first categorize the fields with the same gauge quantum numbers
as follows:
\begin{align}
S \in  \{X_0,N^c\} \ , \qquad
\Phi_u \in \{H_u, X_u\} \ , \qquad
\Phi_d \in \{H_d, X_d, L\} \ .
\end{align}
The operators allowed by the gauge symmetries can be classified
according to their mass dimensions as 
\begin{align}
%\begin{split}
W_2 &= S \ , \\
W_3 &= SW_2 + \Phi_u\Phi_d \ , \\
W_4 &= SW_3 + (Q\Phi_d)D^c + (\Phi_d\Phi_d)E^c + U^c D^c D^c \ , \\
W_5 &= SW_4 + (QQ)(Q\Phi_d) + U^c U^c D^c E^c + (Q\Phi_d)U^c E^c \nonumber \\
         &\quad  + (\Phi_u\Phi_d)(\Phi_u\Phi_d) + (QQ)U^cD^c \ ,
%\end{split}
\end{align}
where 
we suppress coefficients and flavor indices for simplicity. 
We also omit operators involving the adjoint chiral fields
since they have no effect on proton decay.

It is straightforward to see that
all of the operators violating either the baryon or lepton number 
in $W_2$, $W_3$ and $W_4$ are forbidden by the $R$-symmetry:
this includes the usual trilinear terms,
$U^c D^c D^c$, $QD^c L$, $LE^c L$,
and the bilinear terms $H_u L$.
We also note that the mass terms $X_0 N^c$ 
as well as $N^c N^c$ are forbidden by the $U(1)_R$.
Therefore we can define the conserved baryon and lepton numbers
at the renormalizable level.

There are three types of dimension five operators
that violate both the baryon and lepton numbers:
$(QQ)(QL)$, $U^cU^cD^cE^c$ and $N^cU^cD^cD^c$.
Among them, 
the first two are forbidden by the $U(1)_R$,
whereas the last one is $U(1)_R$ symmetric in our flipped assignment.
As a result,
we are left with a unique operator $N^cU^cD^cD^c$,
which is the baryon and lepton number violating operator
in the present model.

As we mentioned in Sec.~\ref{sec:neutrino},
%the Majorana mass terms for the right-handed neutrinos are forbidden
%by our flipped $R$-charge assignment.
%We note that 
$(LH_u)(LH_u)$ operators are allowed, 
but they gives a negligible contribution to neutrino masses
if the cutoff $\LambdaC$ is sufficiently large. 
We also note that
there are dimension five  operators,
$X_0^2 N^cN^c$ and $(X_uX_d) N^c N^c$,
which are $R$-invariant and lepton number violating.
After the $R$-Higgses get the VEVs, these operators
lead to Majorana masses for right-handed neutrinos,
of order $m_{3/2}^2/\Lambda$,
which can safely be neglected 
(unless $m_{3/2}\sim 100\nunits{TeV}$).

%%%%%%%%%%%%%%%%%%%%%%%%%%%%%%%%%%%%%%%%%%%%%%%%%%%%%%%%%%%%%%%%%%%%%%
\section{$U(1)$  Flavor Symmetry and Proton Stability}
\label{sec:FN}
%%%%%%%%%%%%%%%%%%%%%%%%%%%%%%%%%%%%%%%%%%%%%%%%%%%%%%%%%%%%%%%%%%%%%%

As we see in sec.\ref{sec:flipped},
the model is severely constrained from proton decay
if the gravitino is very light, $m_{3/2} < 16~\mathrm{eV}$,
especially for large $\tan\beta$.
In this appendix,
we consider a model with Froggatt-Nielsen $U(1)$ flavor symmetry
to relax the constraint.

Let $\Theta$ be a Froggatt-Nielsen (FN) field 
with $U(1)_F$ charge $-1$,
which is assumed to develop a nonzero VEV $\VEV{\Theta}$.
%slightly below the cutoff $\LambdaC$ of the theory,
We also denote the $U(1)_F$ charge of the fields $\Phi_i$
by the lower case letter $\phi_i$.
The Yukawa terms in \Eq{eq:yukawa}
should be multiplied by a suitable power of the FN field $\Theta$;
for instance, the neutrino Yukawa term
is generated from the $U(1)_F$-invariant operator
\ba
W
\ =\ f_N^{ij}
     \left(\frac{\Theta}{\LambdaC}\right)^{n_i^c + \ell_j + X_u}
     N^c_i L_j X_u 
     \ ,
\non\ea
where $f_N^{ij}$ is a coefficient of order $1$
and $\LambdaC$ is the cutoff.
After we substitute the FN field by its VEV, 
we have $y_N^{ij}=f_N^{ij}\cabbibo^{\nu_i^c + \ell_j + X_u}$.
We will identify $\cabbibo \equiv \VEV{\Theta}/\LambdaC$ 
with the Wolfenstein parameter $\cabbibo=0.22$.
Similarly, we have
%we expect the following suppression of Yukawa couplings
\begin{eqnarray}
  y_U^{ij} = f_U^{ij} \cabbibo^{q_i + u^c_j + h_u}
      \ , \quad 
  y_D^{ij} = f_D^{ij} \cabbibo^{q_i + d^c_j + h_d}
      \ , \quad 
  y_E^{ij} = f_E^{ij}  \cabbibo^{e^c_i + l_j + h_d}
      \ .
 \end{eqnarray}
%where we define $\cabbibo$ as $\cabbibo \equiv \VEV{\Theta}/\LambdaC$.
%\frac{\langle X \rangle}{M_\text{pl}}$ with $M_\text{pl}$ plank scale, 

%From quark and lepton mass hierarchy,
%We can think of following class of  charge assignment\cite{Kakizaki:2002hs} which is given by
%We shall determine the $U(1)$ charge assignment from following fact:

The $U(1)_F$ charge assignment
can be restricted by the following requirements.
%we follow Ref.~\cite{Kakizaki:2002hs}:
First,
%\begin{itemize}
%\item
from the intergenerational mixings of quarks,
$V_{us} \sim \cabbibo$, 
$V_{cb} \sim \cabbibo^2$ and 
$V_{ub} \sim \cabbibo^3$,
%\begin{align}
%V_{us} \sim \cabbibo
%      \ , \quad 
%V_{cb} \sim \cabbibo^2
%      \ , \quad 
%V_{ub} \sim \cabbibo^3
%      \ ,
%\non
%\end{align}
we obtain the following relations
\begin{align}
q_1 = q_3 + 3
      \ , \qquad 
q_2 = q_3 + 2
      \ .
\label{eq:CKM-charge}
\end{align}
%\item
Next,
%quark and charged lepton mass hierarchy can be parameterized as
the quark mass hierarchy can be parameterized as
$m_u : m_c : m_t     \sim \cabbibo^8 : \cabbibo^4 : 1$ and 
$m_d : m_s : m_b     \sim \cabbibo^4 : \cabbibo^2 : 1$,
while the mass hierarchy of charged leptons is
$m_e : m_\mu : m_\tau \sim \cabbibo^5 : \cabbibo^2 : 1$.
%\begin{align}
%       m_u : m_c : m_t     &\sim \cabbibo^8 : \cabbibo^4 : 1
%       \ , 
%\non\\
%       m_d : m_s : m_b     &\sim \cabbibo^4 : \cabbibo^2 : 1
%       \ ,
%\non\\
%       m_e : m_\mu : m_\tau &\sim \cabbibo^5 : \cabbibo^2 : 1
%       \ ,
%\non
%\end{align}
These mass hierarchies can be reproduced, for instance, if we take
%\begin{alignat}{2}
%    q_1 + u^c_1 &=    q_3 + u^c_3 + 8 \ , \qquad 
%    q_2 + u^c_2 &=    q_3 + u^c_3 + 4 \ , \nonumber  \\ 
%    q_1 + d^c_1 &=    q_3 + d^c_3 + 4 \ , \qquad 
%    q_2 + d^c_2 &=    q_3 + d^c_3 + 2 \ , \nonumber \\ 
% \ell_1 + e^c_1 &= \ell_3 + e^c_3 + 5 \ , \qquad 
% \ell_2 + e^c_2 &= \ell_3 + e^c_3 + 2 \ . 
%%\label{eq:mass-charge}
%  %q_3 + d^c_3 &=& e^c_3 + l_3, \quad q_3 + u^c_3 + h_u = 0.\label{charge}
%\non
%\end{alignat}
%Or equivalently,
%\begin{alignat}{3}
%    u^c_1 &=    u^c_3 + 5 \ , \qquad\qquad\qquad
%    u^c_2 &=&   u^c_3 + 2 \ , \nonumber  \\ 
%    d^c_1 &=    d^c_3 + 1 \ , \qquad\qquad\qquad
%    d^c_2 &=&   d^c_3     \ , \nonumber \\ 
% \ell_1 + e^c_1 &= \ell_3 + e^c_3 + 5 \ , \qquad 
% \ell_2 + e^c_2 &=& \ell_3 + e^c_3 + 2 \ . 
%\label{eq:mass-charge}
%  %q_3 + d^c_3 &=& e^c_3 + l_3, \quad q_3 + u^c_3 + h_u = 0.\label{charge}
%\end{alignat}
\begin{alignat}{3}
    u^c_1 &=    u^c_3 + 5 \ , \qquad
    d^c_1 &=    d^c_3 + 1 \ , \qquad
 \ell_1 + e^c_1 &= \ell_3 + e^c_3 + 5 \ , \qquad 
\non\\
    u^c_2 &=   u^c_3 + 2 \ , \qquad 
    d^c_2 &=   d^c_3     \ , \qquad \qquad 
 \ell_2 + e^c_2 &= \ell_3 + e^c_3 + 2 \ . 
\label{eq:mass-charge}
  %q_3 + d^c_3 &=& e^c_3 + l_3, \quad q_3 + u^c_3 + h_u = 0.\label{charge}
\end{alignat}
%
%\item 
The FN charges of the third generation of quarks and leptons
can be constrained,
if we require 
the top Yukawa coupling of order unity, $y_t \sim 1$,
and also the bottom-tau unification, $m_b \sim m_\tau$,
%and null charge for the supersmu terms dictate us to have 
%
which gives
\begin{align}
 q_3 + u^c_3 + h_u = 0 \ , \qquad 
 q_3 + d^c_3 = e^c_3 + \ell_3 \ ,  
\label{eq:FN:third}
\end{align}
%\item 
In addition,
we require for definiteness that
the supersymmetric Higgs mass terms have null charge,
\begin{align}
 x_d + h_u = 0 \ , \qquad  
 x_u + h_d = 0 \ . 
\label{eq:FN:mu}
\end{align}
%\end{itemize}
%Plus, from the fact that bottom quark and tau lepton have approximately same mass,and top Yukawa is of order unity, we get
Note that
the FN charges of $H_u$ and $H_d$
can be chosen independently
since there is no $\mu H_u H_d$ term in the present $U(1)_R$ symmetric model.

%Only the difference between the MSSM case \cite{Kakizaki:2002hs} 
%and the present choice
%is the charge relation for Higgs in \Eq{eq:FN:mu}
%and that we have right handed neutrino $N_i^c$. 
%whose Yukawa couplings are controlled by the gravitino mass.
%If we apply Froggatt-Nielsen mechanism to neutrino Yukawa coupling,
%we can expect following relation

Now,
we discuss the suppression of the dimension five operator \eq{eq:dim.5}
by the FN mechanism.
The coefficient $C_5=C_5^{1123}$ of the operator 
$U^c_1 D^c_1 D^c_2 N^c_3 $
receives a suppression of 
$C_5 \sim \cabbibo^{u^c_1 + d^c_1 + d^c_2 + n^c_3}$.
The FN charge $n^c_3$ of the right-handed neutrino
can be eliminated 
if we use the relation \eq{eq:Ynu:simplify}
\begin{equation}
 y_\nu^{ij}
  =   f_\nu^{ij}  
      \cabbibo^{n^c_i + l_j + x_u}
  = \frac{m^{ij}_\nu}{m_{3/2}}
      \left(\frac{\mu_2^2 + m^2_{X_u}}{\mu_2 v\cos\beta}\right)
      \ .
\label{eq:ynu}
\end{equation}
Combining \Eqs{eq:CKM-charge}{eq:FN:mu} with \Eq{eq:ynu},
we obtain
\begin{align}
C_5 
 &=  C_5^{(0)}
      \cabbibo^{u^c_1 + d^c_1 + d^c_2 + n^c} 
  = C_5^{(0)}
      \cabbibo^{6+Q}
      \left(\frac{m_\nu}{m_{3/2}}\right)
      \left(\frac{\mu_2^2 + m^2_{X_u}}{\mu_2 v\cos\beta}\right)
      \ ,
\label{eq:dim5:suppressed}
\end{align}
where 
\begin{align}
Q   &\equiv 
    h_d - h_u + d^c_3 + e^c_3 -2q_3
    \ .
\end{align}
We see that
the coefficient $C_5$ %of the operator $N^c U^c_1 D^c_1 D^c_2$
is suppressed for several reasons:
the suppression factor $\cabbibo^6$
comes from the FN charges of the 1st and 2nd generations
of quarks involved in the proton decay.
Another factor $\cabbibo^Q$
expresses the dependence on the FN charges of the Higgses
and the 3rd generation of quarks and leptons.
For instance,
if we assign a negative FN charge to the MSSM Higgs $H_u$,
the up-type quarks have positive FN charges,
which suppress the dimension five operator.
Notice that
this can be done without suppressing the supersymmetric mass terms
since we can assign the opposite FN charge to the $R$-partner $X_d$.

It should be noticed that
there is yet another suppression in \Eq{eq:dim5:suppressed}:
the coefficient $C_5$ is more suppressed for larger gravitino mass.
%It follows that
%the proton lifetime is proportional to
%the fourth power of $m_{3/2}$. instead of .......
This dependence can be understood as follows.
Assigning a positive FN charge $n^c>0$ to the right-handed neutrinos
suppresses 
not only the coefficient $C_5$ of the dimension five operator
but also the neutrino Yukawa coupling $y_{\nu}$;
$y_\nu\sim\cabbibo^{n^c}$ and 
$C_5\sim\cabbibo^{n^c}$,
so that the proton lifetime scales as
$\tau_p\sim\cabbibo^{-4n^c}$.
For a fixed value of the neutrino mass,
the suppression of the neutrino Yukawa coupling
is translated into the dependence on $m_{3/2}$ in \Eq{eq:dim5:suppressed}.

%Fig.~\ref{fig:FN} shows how the constraint
%from the proton decay can be satisfied
%for a moderate choice of the FN charge $Q$.
%We note that
%the gravitino mass smaller than $\order{10\nunits{eV}}$
%can satisfy the proton stability
%thanks to the flavor symmetry.

\begin{figure}[tb]
\begin{center}
\includegraphics[width=70mm]{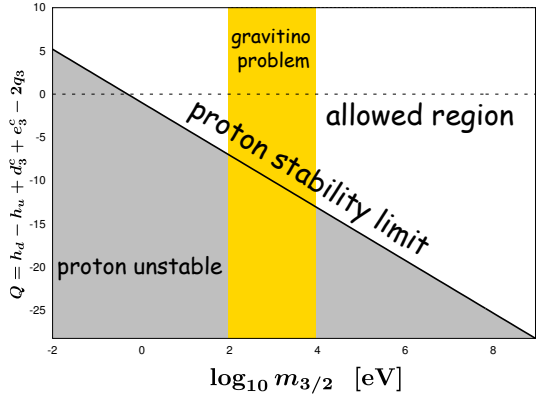}
\caption{
The constraint on the FN charge $Q$.
%Super kamiokande limit line.
%The upper line represent cutoff scale 
%as a function of gravitino mass 
%and the lower line corresponds to the case with $U(1)_F$ flavor symmetry.
The parameters are the same as in Fig.~\ref{fig:protonstability}.
%$\tan\beta = 3$,
%$\mu_2 = m_{X_u}= 300\nunits{GeV}$,
%and $m_{\wt{q}} \sim 1\nunits{TeV}$
}
\label{fig:FN}
\end{center}
\end{figure}

Putting it all together,
a similar analysis as in Sec.~\ref{sec:protondecay}
gives an estimate of the proton lifetime as
\begin{align}
\tau_p 
 = \frac{3.7\times10^{34}\nunits{yr}}{%C_5^{(0)2}
   \cabbibo^{2Q}}
    \left(\frac{\LambdaC}{2.4 \times 10^{18}\nunits{GeV}}\right)^2
    \left(\frac{m_{3/2}}{1 \nunits{eV}}\right)^4
    \left(\frac{\cos^2\beta}{0.1}\right)^3
 \ ,
\end{align}
where we have set the coefficient to be $C_5^{(0)}=1$.
Fig.~\ref{fig:FN} shows how the constraint
from the proton decay can be satisfied
for a moderate choice of the FN charge $Q$.
We see that 
the proton decay constraint can be satisfied
even for $m_{3/2}=1\nunits{eV}$ 
if we take $Q=0$ and $\tan\beta=3$.
In this way,
the present model combined with the $U(1)_F$ flavor symmetry evades
the proton decay constraint in a wider region of parameter space.

The situation is quite different
from that in the MSSM case \cite{Kakizaki:2002hs},
in which a negative FN charge of the Higgses
would imply unacceptably small $\mu$ term.
In fact, 
the analysis in Ref.~\cite{Kakizaki:2002hs} shows
that
the FN charge of the $\mu$ term,
\ie, the sum of the FN charges of $H_u$ and $H_d$,
are tightly constrained
since both of LLLL and RRRR operators \eq{eq:dim5:MSSM}
should be suppressed simultaneously.
In our case,
we have 
only RRRR type operator
since LLLL operator is forbidden by the $U(1)_R$ symmetry.

%the proton decay can be suppressed
%adopting the Froggatt--Nielsen mechanism
%in two different ways:
%\begin{itemize}
%\item
%a
%\item
%a
%Visible SUSY breaking mode, on the other hand,
%need not to have $\mu H_u H_d$ term from the first place.
%\end{itemize}

%\input{sec_operator}
%\input{sec_FN}

%%%%%%%%%%%%%%%%%%%%%%%%%%%%%%%%%%%%%%%%%%%%%%%%%%%%%%%%%%%%%%%%%%%%%%
% Bibliography

\bibliographystyle{apsrev}
\bibliography{biblio}

%%%%%%%%%%%%%%%%%%%%%%%%%%%%%%%%%%%%%%%%%%%%%%%%%%%%%%%%%%%%%%%%%%%%%%
\end{document}